\documentclass[twocolumn,superscriptaddress,nofootinbib,amsmath,amssymb,aps,pra,floatfix]{revtex4-1}

\usepackage{comment}
\usepackage[utf8]{inputenc}
\usepackage{wrapfig}
\usepackage{float}
\usepackage{amssymb}
\usepackage{amsfonts}
\usepackage{amsmath}
\usepackage{cases}
\usepackage{stmaryrd}\usepackage{tipa}
\usepackage[dvips]{graphicx}
\usepackage{dcolumn}
\usepackage{bbold}
\usepackage{graphicx}
\usepackage{graphicx}
\usepackage{subfigure}
\usepackage{dcolumn}
\usepackage{bm}
\usepackage{color}
\usepackage{xcolor}
\usepackage{subfigure}
\usepackage{physics}
\usepackage{array}
\usepackage{multirow}
\usepackage{nopageno}
\usepackage{mathtools}
\usepackage{enumerate}
\usepackage[colorlinks=true, linkcolor=red, citecolor=blue]{hyperref}
\usepackage{diagbox}
\usepackage{ulem}
\usepackage{diagbox}
\usepackage{makecell}

\begin{document}

\footnotetext{This manuscript has been authored by UT-Battelle, LLC, under Contract No. DE-AC0500OR22725 with the U.S. Department of Energy. The United States Government retains and the publisher, by accepting the article for publication, acknowledges that the United States Government retains a non-exclusive, paid-up, irrevocable, world-wide license to publish or reproduce the published form of this manuscript, or allow others to do so, for the United States Government purposes. The Department of Energy will provide public access to these results of federally sponsored research in accordance with the DOE Public Access Plan.}

\title{Digital-Analog Quantum Simulations Using The Cross-Resonance Effect}

\author{Tasio Gonzalez-Raya}
\affiliation{Department of Physical Chemistry, University of the Basque Country UPV/EHU, Apartado 644, 48080 Bilbao, Spain}

\author{Rodrigo Asensio-Perea}
\affiliation{Department of Physical Chemistry, University of the Basque Country UPV/EHU, Apartado 644, 48080 Bilbao, Spain}

\author{Ana Martin}
\affiliation{Department of Physical Chemistry, University of the Basque Country UPV/EHU, Apartado 644, 48080 Bilbao, Spain}

\author{Lucas C. C\'{e}leri}
\affiliation{Institute of Physics, Federal University of Goi\'{a}s, 74.690-900 Goi\^{a}nia, Goi\'{a}s, Brazil}

\author{Mikel Sanz}
\affiliation{Department of Physical Chemistry, University of the Basque Country UPV/EHU, Apartado 644, 48080 Bilbao, Spain}
\affiliation{IKERBASQUE, Basque Foundation for Science, Plaza Euskadi 5, 48009 Bilbao, Spain}
\affiliation{IQM, Nymphenburgerstr. 86, 80636 Munich, Germany}

\author{Pavel Lougovski}
\email{\text{Now at Amazon Web Services}}
\affiliation{Quantum Information Science Group, Oak Ridge National Laboratory, Oak Ridge, Tennessee 37831, USA}

\author{Eugene F. Dumitrescu}
\email{dumitrescuef@ornl.gov}
\affiliation{Quantum Information Science Group, Oak Ridge National Laboratory, Oak Ridge, Tennessee 37831, USA}

\begin{abstract}

Digital-analog quantum computation aims to reduce the currently infeasible resource requirements needed for near-term quantum information processing by replacing sequences of one- and two-qubit gates with a unitary transformation generated by the systems' underlying Hamiltonian. Inspired by this paradigm, we consider superconducting architectures and extend the cross-resonance effect, up to first order in perturbation theory, from a two-qubit interaction to an analog Hamiltonian acting on 1D chains and 2D square lattices which,  in an appropriate reference frame, results in a purely two-local Hamiltonian. By augmenting the analog Hamiltonian dynamics with single-qubit gates we show how one may generate a larger variety of distinct analog Hamiltonians. We then synthesize unitary sequences, in which we toggle between the various analog Hamiltonians as needed, simulating the dynamics of Ising, $XY$, and Heisenberg spin models. Our dynamics simulations are Trotter error-free for the Ising and $XY$ models in 1D. We also show that the Trotter errors for 2D $XY$ and 1D Heisenberg chains are reduced, with respect to a digital decomposition, by a constant factor. In order to realize these important near-term speedups, we discuss the practical considerations needed to accurately characterize and calibrate our analog Hamiltonians for use in quantum simulations. We conclude with a discussion of how the Hamiltonian toggling techniques could be extended to derive new analog Hamiltonians which may be of use in more complex digital-analog quantum simulations for various models of interacting spins. 
\end{abstract}

\maketitle
\section{Introduction}
Classical computers are ill-suited for simulating quantum systems due to their exponentially growing Hilbert spaces. Feynman~\cite{Feynman1982} therefore suggested that it would be more efficient to simulate a quantum system using other, controllable, quantum systems. This idea gave birth to the research area of quantum simulation~\cite{Georgescu2014}. 

The simulation of purely quantum features, such as entanglement and superposition, is very costly to represent on classical computers, whereas on a quantum system these features arise naturally. A quantum simulator is a quantum platform, such as trapped ions~\cite{Blatt2012} or cold atoms~\cite{Bloch2012}, over which we have great controllability. Simulators are typically categorized as either digital or analog. An analog simulator makes use of the simulator's underlying Hamiltonian in order to mimic the target system's dynamics, whereas a digital simulator approximates the target system's Hamiltonian evolution through a composition of one- and two-qubit gates drawn from a universal gate set. Nevertheless, there are other possible realizations of quantum simulators. A quantum annealer uses quantum fluctuations to efficiently solve optimization problems, but it can also be used as an adiabatic quantum simulator~\cite{Babbush2015,Roth2019}. 

Going beyond this distinction, a novel paradigm for digital-analog (DA) quantum computation~\cite{Parra2020,Martin2020,Galicia2020,Headley2020} and simulation~\cite{Mezzacapo2014,Yung2014,Arrazola2016,Lamata2018,Kyriienko2018,Hegade2020} has been proposed. These DA schemes combine the application of fast digital single-qubit gates with multi-qubit interactions provided by an underlying analog Hamiltonian~\cite{Dodd2002}. Leveraging the natural interaction between qubits as an analog resource, DA schemes for the quantum approximate optimization algorithms and the quantum Fourier transform have been shown to be more error resilient, especially as the size of the simulation scales up~\cite{Martin2020,Headley2020}. Therefore, the DA quantum computation paradigm provides an attractive near-term solution to alleviate the current difficulties associated with implementing useful quantum algorithms with near term devices. Despite this promise, the success of the DA approach relies on having a quantum platform with well-defined qubits, controllable pulses, and an accurate characterization of the underlying interaction Hamiltonian.

At the moment, superconducting circuits have been established as a leading quantum platform in terms of controllability and scalability, mainly caused by the introduction of the transmon qubit~\cite{Koch2007}. Implementations controlled by microwave pulses have achieved very low errors on single-qubit gates~\cite{McKay2017}, and the most common two-qubit gate for fixed frequency transmons is based on the cross-resonance (CR) interaction~\cite{Paraoanu2006,Rigetti2010,Chow2011}. The CR gate uses a single microwave pulse to entangle a pair of fixed-frequency qubits, making use of a static coupling. Despite some success, constructing high-fidelity controlled-NOT operations with the CR gate in multi-qubit devices remains a field of active research~\cite{Sheldon2016,Magesan2020,Malekakhlagh2020,Sundaresan2020,Ku2020}.

In this article, we consider a CR gate interaction between two superconducting qubits in order to obtain a purely non-local, in a particular frame, effective interaction Hamiltonian. Further, we consider a multi-qubit extension and derive the generalized effective multi-qubit two-local Hamiltonian. Next, we consider how the multi-qubit Hamiltonian may be toggled into a variety of forms using digital single qubit gates. Utilizing the resulting set of Hamiltonians we design DA protocols to simulate Ising, $XY$, and Heisenberg spin models. The resulting DA sequences are in some cases Trotter-error free in 1D. We compute the Trotter error when it is present and find that it is reduced by a constant factor with respect to a Digital decomposition of the same model.

\begin{figure}[t]
{\includegraphics[width=0.5 \textwidth]{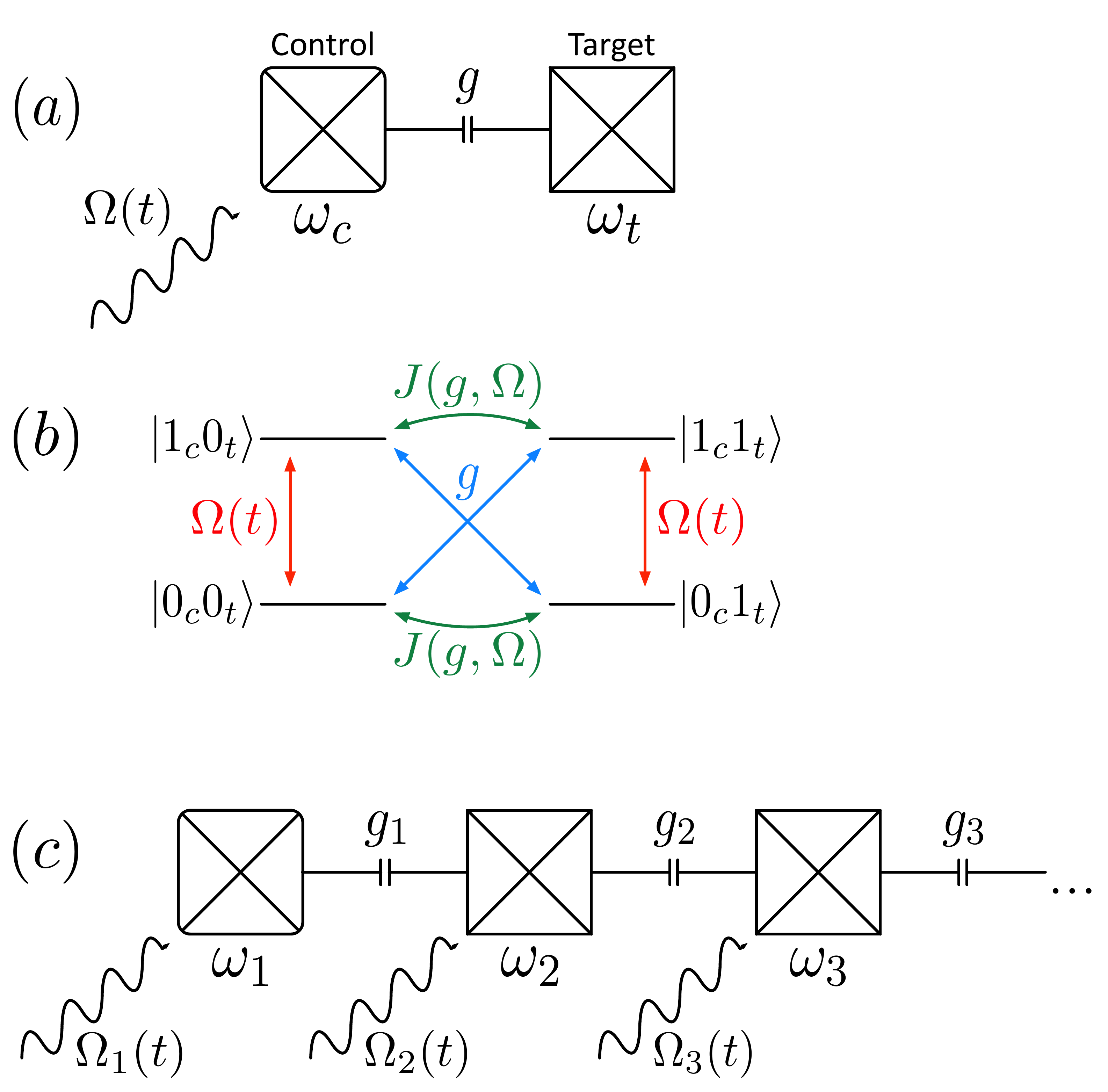}}
\caption{Graphical representation of the cross-resonance effect: (a) Two qubits, the first one being the control qubit with resonance frequency $\omega_{c}$ and the second one the target qubit with resonance frequency $\omega_{t}$, are interacting with strength $g$. The control qubit is driven at the resonance frequency of the target qubit, with driving amplitude $\Omega(t)$. (b) State space representation of the transitions between levels of the control and target qubits, in the presence of a driving of amplitude $\Omega(t)$ on the control qubit. The effective cross-resonance interaction is described by strength $J(g,\Omega)$. (c) $N$ qubits with nearest neighbour interaction, all of them are driven at the resonance frequency of their neighbour to the right, illustrating the scenario we describe in Sec.~\ref{sec:N_qubits}.}
\label{fig1}
\end{figure}

\section{Deriving the effective Cross-Resonance Hamiltonian}
\label{sec:CRHamiltonians}

In this section we present the effective CR Hamiltonians, derived in the manner described in Ref.~\cite{Rigetti2010}. We first introduce the two-qubit scenario, in order to develop an intuition for the effective coupling, and then generalize the results to the case of $N$ qubits. Further details of the calculations, supporting the main text, can be found in Appendix~\ref{app:CRHamiltonians}. Note that, in this manuscript, we are working with $\hbar =1$.

\subsection{Two qubits}
Our starting point is the laboratory frame Hamiltonian, written as
\begin{eqnarray}
\nonumber H_{LAB} &=& \frac{1}{2}(\omega_{1}^{q}z_{1} +\omega_{2}^{q}z_{2}) + \Omega_{1}x_{1}\cos(\omega_{1} t+\phi_{1}) \\
&& + \Omega_{2}x_{2}\cos(\omega_{2} t+\phi_{2}) + \frac{g}{2}x_{1}x_{2},
\label{eq:Ham2Q}
\end{eqnarray}
where $x_i, y_i, z_i$ are the Pauli matrices supported on site $i$, $\omega_{k}^{q}$ and $\omega_{k}$ are the resonance and the driving frequencies of qubit $k$, respectively. $\Omega_{k}$ represents the amplitude of the driving field, while $g$ denotes the strength of the interaction between the qubits. 

The effective Hamiltonian is derived by applying a series of unitary transformations --- described in detail in Appendix~\ref{sec:2_qubit_details} --- to Eq.~\ref{eq:Ham2Q}. First, we apply a double rotation into the frame co-rotating at the driving frequency of the qubits ($\omega_1, \omega_2$). After this, we apply the rotating wave approximation (RWA), valid for  $\omega_1,\omega_2 \gg \delta_i = \omega_{1}^{q} - \omega_{1}, \Omega_i, g$, to drop fast terms rotating with frequency $\pm 2 \omega_1, \pm 2 \omega_2, \pm (\omega_1+\omega_2)$. We then proceed by applying two new rotations in order to express the Hamiltonian in a more convenient frame, named the quad frame (QF). In this frame, all local terms are eliminated and the result is a purely two-local Hamiltonian. The next step is to consider the case in which we drive the first qubit at the resonance frequency of the second qubit, $\omega_1 = \omega_{2}^{q}$, while the second one is not driven, as can be seen in Fig.~\ref{fig1}(a). After a final RWA, valid for $\Omega_1 \gg g$ or $\delta \gg g$, we end up with the effective Hamiltonian
\begin{equation}\label{eq:H_QF}
H_{QF} = \frac{g\Omega_{1}}{4\delta}(\cos\phi_{1}x_{1}x_{2} + \sin\phi_{1}x_{1}y_{2}).
\end{equation}
As $\phi_{1}$ is a controllable phase, we can set $\phi_{1}=0$, resulting in
\begin{equation}\label{H_eff_QF_2Q}
H_{QF} = \frac{g\Omega_{1}}{4\delta} x_{1}x_{2}.
\end{equation}

\subsection{$N$ qubits}
\label{sec:N_qubits}
The $N$-qubit Hamiltonian, in the laboratory frame, is given by
\begin{equation}
H_{LAB} = \sum_{k=1}^{N}\left[ \frac{\omega^{q}_{k}}{2}z_{k} + \Omega_{k}x_{k}\cos(\omega_{k}t+\phi_{k}) \right] +\sum_{k=1}^{N-1}\frac{g_{k}}{2}x_{k}x_{k+1}
\label{eq:HamilNq}
\end{equation}
We proceed by moving to the QF by means of appropriate rotations (see Appendix~\ref{sec:N_qubit_details} for details). The driving field is then applied to all qubits at the resonance frequency of their neighbour to the right, as shown in Fig.~\ref{fig1}(c), except for case of open boundary conditions in which case the last qubit is not driven. 
Similar to the two qubit case, the frame transformations re-express the Hamiltonian in a purely two-local form. Keeping only terms linear in $\Omega_i/\delta_i$, and neglecting fast oscillating terms $\delta \gg g$ by RWA, we arrive at the effective Hamiltonian 
\begin{eqnarray}\label{H_eff_QF_complete}
\nonumber H_{QF} &=& \sum_{k=1}^{N-1}\frac{g_{k}\Omega_{k}}{4\delta_{k}}x_{k}(y_{k+1}\sin(\phi_{k}-\phi_{k+1}) \\
&-& z_{k+1}\cos(\phi_{k}-\phi_{k+1})).
\end{eqnarray}
Once again, we have the freedom to set $\phi_{k}=\phi$ for all k. The Hamiltonian then reduces to
\begin{equation}\label{H_QF_eff}
H_{QF} = \sum_{k=1}^{N-1}J_{k}x_{k}z_{k+1},
\end{equation}
where we have defined $J_{k}=-g_{k}\Omega_{k}/4\delta_{k}$. As seen in the two-qubit case, the Hamiltonian only contains two-qubit interaction terms. In the next sections we will discuss the use of this Hamiltonian to generate the analog dynamics of a DA computation. 


\section{Digital-Analog Computing}
We take Eq.~\ref{H_QF_eff} as a starting point, and consider $\Omega_{k} = \Omega$, $\delta_{k}=\delta$, $g_{k} = g$, $J_{k}=J$, for simplicity. Then, we write the effective Hamiltonian in the QF as
\begin{equation}\label{H_control}
H_{A} = J\sum_{k=1}^{N-1}x_{k}z_{k+1}.
\end{equation}



\subsection{Synthesis Error}
Given that the effective Hamiltonian is the center piece of the simulation protocols, we need to estimate the synthesis error associated to the fact that it is an approximation of the original Hamiltonian. In the weak-driving regime $\Omega_{k} \ll \delta_{k}$, the original Hamiltonian without the QF RWA is
\begin{eqnarray}
H^{org} &=& \frac{g}{4} \sum_{k=1}^{N-1} \bigg\{ (z_{k}z_{k+1} + y_{k}y_{k+1})\cos\delta t \\
\nonumber &&+ (y_{k}z_{k+1} - z_{k}y_{k+1})\sin\delta t \\
\nonumber && -\frac{\Omega}{\delta} \Big[ x_{k}z_{k+1} + (z_{k}\cos2\delta t + y_{k}\sin2\delta t)x_{k+1}\Big] \bigg\}.
\end{eqnarray}
In order to compute the synthesis error, we focus on the Frobenius norm, 
\begin{equation}
|| A ||_{F} = \sqrt{\tr(A^{\dagger}A)},    
\end{equation}
which provides an upper bound for the spectral norm. Let us compute the norm for the difference between the two Hamiltonians, $\Delta H = H^{org}-H_{A}$,
\begin{eqnarray}
\Delta H &=& \frac{g}{4} \sum_{k=1}^{N-1} \bigg\{ (z_{k}z_{k+1} + y_{k}y_{k+1})\cos\delta t \\
\nonumber && + (y_{k}z_{k+1} - z_{k}y_{k+1})\sin\delta t  \\
\nonumber && -\frac{\Omega}{\delta} (z_{k}\cos2\delta t + y_{k}\sin2\delta t)x_{k+1} \bigg\}.
\end{eqnarray}
The latter part of this operator contributes with $\Omega^{2}/\delta^{2}$ to the Frobenius norm, so we will neglect that part in the approximation $\Omega/\delta \ll 1$. The rest can be written as
\begin{eqnarray}
\Delta H &=& \frac{g}{4} \sum_{k=1}^{N-1} \bigg\{ (z_{k}\cos\delta t + y_{k}\sin\delta t) z_{k+1} \\
\nonumber && + (y_{k}\cos\delta t - z_{k}\sin\delta t) y_{k+1} \bigg\},
\end{eqnarray}
which corresponds to the result of a rotation given by $U_{k} = e^{-i\delta t x_{k}/2}$. This norm can be computed analytically by rewriting the last expression as
\begin{equation}
\Delta H = \frac{g}{4}\sum_{k=1}^{N-1}U_{k}^{\dagger}(z_{k}z_{k+1}+y_{k}y_{k+1})U_{k}.
\end{equation}
Then, we see that the only terms that survive the trace of
\begin{eqnarray}
&& (\Delta H)^{\dagger}\Delta H = \frac{g^{2}}{16} \times \\
\nonumber && \sum_{k,k'=1}^{N-1}U_{k}^{\dagger}(z_{k}z_{k+1}+y_{k}y_{k+1})U_{k}U_{k'}^{\dagger}(z_{k'}z_{k'+1}+y_{k'}y_{k'+1})U_{k'}
\end{eqnarray}
are those which satisfy $k=k'$. Consequently, we obtain
\begin{equation}
\tr[(\Delta H)^{\dagger}\Delta H] = \frac{g^{2}}{16}\tr(2\sum_{k=1}^{N-1}\mathbb{1}) = \frac{g^{2}}{8}(N-1)\tr(\mathbb{1}),
\end{equation}
where $\mathbb{1}$ actually represents $\bigotimes_{k=1}^{N} \mathbb{1}_{k}$. We want to set the normalization to $\tr(\mathbb{1})=1$, which corresponds to a factor of $2^{-N/2}$ on the Frobenius norm, since
\begin{equation}
\left|\left| \bigotimes_{k=1}^{N} \mathbb{1}_{k} \right|\right|_{F} = 2^{N/2}.
\end{equation}
Then, we find the Frobenius norm for N qubits ($N\geq 2$) to be
\begin{equation}
|| \Delta H ||_{F} = \frac{g}{2\sqrt{2}} \sqrt{N-1}.
\end{equation}
See that this norm diverges with the square root of the number of qubits. Notice however that the Frobenius norm per qubit decreases with $N$. Furthermore, we have computed the norm of the difference between the propagators, $\Delta \hat{P} = \hat{P}^{org}-\hat{P}_{A}$,
\begin{equation}
|| \Delta \hat{P} ||_{F} = \frac{g}{\delta\sqrt{2}} \left| \sin\frac{\delta t}{2} \right| \sqrt{N-1}.
\end{equation}
Here, the propagators are computed up to first order in the Dyson series. Again, the norm of the difference of propagators per qubit decreases with $N$. Note that, for $\delta t \ll 1$, 
\begin{equation}
|| \Delta \hat{P} ||_{F} \approx t \cdot || \Delta H ||_{F}. 
\end{equation}

The synthesis errors corresponding to the Hamiltonians derived in further sections can be found in the Appendix~\ref{app:synthesiserror}.


\subsection{Hamiltonian toggling}

Let us now consider DA quantum simulations of the spin-1/2 Ising, $XY$, and Heisenberg models in 1 and 2 dimensions. We designate the effective Hamiltonian in the QF, given in Eq.~\ref{H_QF_eff}, as our fundamental DA Hamiltonian from which all others will be generated. Rotating to the reference frame where the Hadamard transformation is applied to all {\em even} qubits, i.e. $U^{e}=\bigotimes_i W_{2i}$, the Hamiltonian transforms into 
\begin{equation}\label{H_even}
H^e= J\sum_{k=1}^{\frac{N}{2}}x_{2k-1}x_{2k} + J\sum_{k=1}^{\frac{N-1}{2}}z_{2k} z_{2k+1}. 
\end{equation}
From this reference frame, Hadamard transforming all qubits will toggle the Hamiltonian into its odd form, i.e. translating the Hamiltonian by one site,
\begin{equation}\label{H_odd}
H^o= J\sum_{k=1}^{\frac{N}{2}}z_{2k-1}z_{2k} + J\sum_{k=1}^{\frac{N-1}{2}}x_{2k} x_{2k+1}.  
\end{equation}

\subsection{Two-dimensional generalization}
\begin{figure}
    \centering
    \includegraphics[width=\columnwidth]{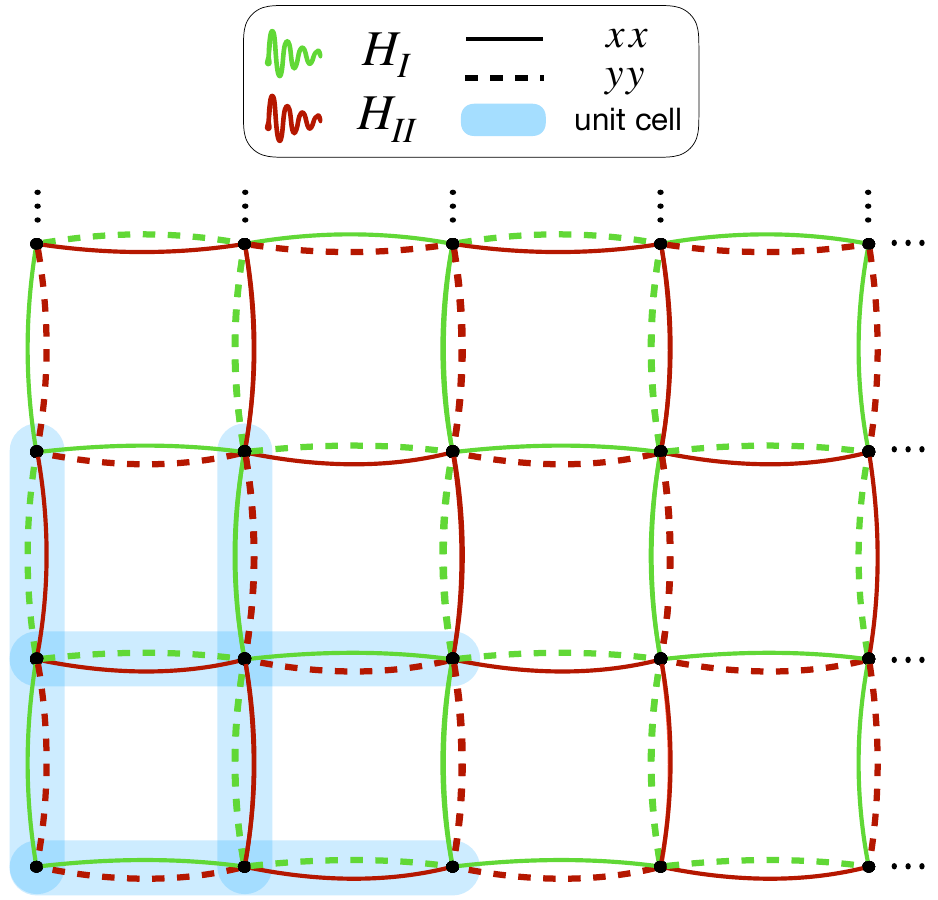}
    \caption{Illustration of analog Hamiltonian interactions on a 2-dimensional lattice. The green (red) lattice represents the Hamiltonian given by Eq.~\ref{eq:H_{I}} (\ref{eq:H_{II}}). Vertices correspond to qubits in a 2D lattice and the solid and dashed edges correspond to the $xx$ and $yy$ interactions, respectively.}
    \label{fig2}
\end{figure}

Let us also consider the extension of the Hamiltonian to two dimensions. Consider a single target qubit in a two dimensional lattice which is driven at the frequencies of its neighbors in the $+\hat{i}$ and $+\hat{j}$ directions. This realizes a $x_c z_t$-interaction between the control qubit located at $(i,j)$ and target qubits at sites $(i+1,j)$ and $(i,j+1)$. The extension of $H^{o}$ in Eq.~\ref{H_odd} is
\begin{eqnarray}
\nonumber H_{2D}^{o} &=& J \bigg[ \sum_{i=1}^{\frac{N}{2}}\sum_{j=1}^{\frac{N}{2}} z_{2i-1,2j-1}(z_{2i-1,2j}+z_{2i,2j-1}) \\
\nonumber &+& \sum_{i=1}^{\frac{N-1}{2}}\sum_{j=1}^{\frac{N-1}{2}} z_{2i,2j}(z_{2i,2j+1}+z_{2i+1,2j}) \\
\nonumber &+& \sum_{i=1}^{\frac{N}{2}}\sum_{j=1}^{\frac{N-1}{2}} x_{2i-1,2j}(x_{2i-1,2j+1}+x_{2i,2j}) \\
&+& \sum_{i=1}^{\frac{N-1}{2}}\sum_{j=1}^{\frac{N}{2}} x_{2i,2j-1}(x_{2i,2j}+x_{2i+1,2j-1}) \bigg],
\end{eqnarray}
where summations run over repetitions of the unit cell illustrated in Fig.~\ref{fig2}. Likewise, the extension of $H^{e}$ in Eq.~\ref{H_even} is $H_{2D}^{e} = H_{2D}^{o}(x\leftrightarrow z)$, which is easily realized by applying a Hadamard on each site of the lattice. Applying a global $R_x(\pi/2)=e^{-i\pi x/4}$ transformation on Hamiltonian $H_{2D}^{e}$, we obtain
\begin{eqnarray}\label{eq:H_{I}}
\nonumber H_{I} &=& J\sum_{i,j=1}^{\frac{N}{2}} (x_{2i-1,2j-1} x_{2i,2j-1} + y_{2i,2j-1} y_{2i+1,2j-1} \\
\nonumber &+& y_{2i-1,2j} y_{2i,2j} + x_{2i,2j} x_{2i+1,2j} \\
\nonumber &+& x_{2i-1,2j-1} x_{2i-1,2j} + y_{2i,2j-1} y_{2i,2j} \\ 
&+& y_{2i-1,2j} y_{2i-1,2j+1} + x_{2i,2j} x_{2i,2j+1}),
\end{eqnarray}
where we have simplified the summation limits by considering that the Hamiltonian acts on a system with periodic boundary conditions. If we rotate $H_{2D}^{o}$ by $R_x(\pi/2)$, we have
\begin{eqnarray}\label{eq:H_{II}}
\nonumber H_{II} &=& J\sum_{i,j=1}^{\frac{N}{2}} (y_{2i-1,2j-1} y_{2i,2j-1} + x_{2i,2j-1} x_{2i+1,2j-1} \\
\nonumber &+& x_{2i-1,2j} x_{2i,2j} + y_{2i,2j} y_{2i+1,2j} \\
\nonumber &+& y_{2i-1,2j-1} y_{2i-1,2j} + x_{2i,2j-1} x_{2i,2j}  \\ 
&+& x_{2i-1,2j} x_{2i-1,2j+1} + y_{2i,2j} y_{2i,2j+1}).
\end{eqnarray}
Note that $H_{II}$ is just a translation of $H_{I}$ by the vector $(1,1)$. The interactions described by these Hamiltonians are represented in Fig.~\ref{fig2}, where $H_{I}$'s and $H_{II}$'s interactions are illustrated by the green and red edges, respectively. In both cases, the solid (dashed) edges correspond to $xx \: (yy)$ interactions between adjacent qubits, and the summations in Eqs.~\ref{eq:H_{I}}, \ref{eq:H_{II}} correspond to a tiling of the 2D lattice using the unit cell, highlighted in blue in Fig.~\ref{fig2}.


\section{Many-body compilation}
We now discuss how to simulate a variety of paradigmatic spin models with the Hamiltonians discussed above.

\subsection{Ising model}
So far we have considered a multi-qubit framework in which we drive all qubits at the resonance frequency of their neighbours to the right. For this particular case, let us now explore a scenario in which we drive only odd or even qubits, which can be achieved by tuning the system's parameters in the following way:
\begin{eqnarray}
\nonumber k \, \text{control} &\rightarrow& \{ \omega_{k}=\omega_{k+1}^{q}, \varphi_{k}(t) = \delta_{k+1}t + \phi_{k}-\phi_{k+1}, \\
\nonumber && \eta_{k}\approx\delta_{k}, \sin\xi_{k}\approx 1, \cos\xi_{k}\approx\frac{\Omega_{k}}{\delta_{k}}\}, \\
\nonumber k \, \text{target} &\rightarrow& \{ \varphi_{k}(t) = (\omega_{k}-\omega_{k+1})t -\phi_{k+1}, \Omega_{k}=0, \delta_{k}=0, \\
&& \eta_{k}=0, \phi_{k}=0, \sin\xi_{k}=0, \cos\xi_{k}=1\},
\end{eqnarray}
where the qubit we drive is the control qubit and it’s neighbour to the right is the corresponding target qubit. Assuming we drive only odd qubits, the choice of parameters leads to a particular QF transformation, represented by
\begin{equation}
U_{QF}^{\text{odd}} = \bigotimes_{k\,\text{odd}} U_{QF}^{(k)}U_{I}^{(k+1)},
\end{equation}
where $U_{QF}^{(k)}$ is the QF transformation applied on qubit $k$ (this transformation is discussed in Appendix~\ref{appB}), and $U_{I}^{(k+1)}=e^{-\frac{it}{2}\omega^{q}_{k+1}z_{k+1}}$ is the transformation to the interaction picture of qubit $k+1$. After applying a RWA by keeping the static terms, we write the Hamiltonian in the QF as
\begin{equation}
H_{QF}^{\text{odd}} = J \sum_{k=1}^{\frac{N}{2}} x_{2k-1}(x_{2k}\cos\phi + y_{2k}\sin\phi),
\end{equation}
after setting $\delta_{2k-1}=\delta$, $\Omega_{2k-1}=\Omega$, $g_{2k-1}=g$, $\phi_{2k-1}=\phi$, and defining $J=g\Omega/4\delta$. See that this is a straightforward multi-qubit extension of the Hamiltonian in Eq.~\ref{eq:H_QF}. If we do the same, in the case in which we drive only even qubits, the transformation becomes
\begin{equation}
U_{QF}^{\text{even}} = \bigotimes_{k\,\text{even}} U_{QF}^{(k)}U_{I}^{(k+1)},
\end{equation}
and we obtain
\begin{equation}
H_{QF}^{\text{even}} = J \sum_{k=1}^{\frac{N-1}{2}} x_{2k}(x_{2k+1}\cos\phi + y_{2k+1}\sin\phi).
\end{equation}
Considering $\phi=0$, these Hamiltonians become
\begin{eqnarray}
\nonumber H_{QF}^{\text{odd}} &=& J \sum_{k=1}^{\frac{N}{2}} x_{2k-1}x_{2k}, \\
H_{QF}^{\text{even}} &=& J \sum_{k=1}^{\frac{N-1}{2}} x_{2k}x_{2k+1},
\end{eqnarray}
and we see that $[H_{QF}^{\text{odd}},H_{QF}^{\text{even}}]=0$. If we rotate all qubits by a Hadamard gate, we obtain
\begin{eqnarray}
\nonumber U^{\dagger}_{W}H_{QF}^{\text{odd}}U_{W} &=& J \sum_{k=1}^{\frac{N}{2}} z_{2k-1}z_{2k} \equiv H_{1}, \\
U^{\dagger}_{W}H_{QF}^{\text{even}}U_{W} &=& J \sum_{k=1}^{\frac{N-1}{2}} z_{2k}z_{2k+1} \equiv H_{2},
\end{eqnarray}
which leads to
\begin{equation}
H_{ZZ} = H_{1}+H_{2} = J\sum_{k=1}^{N-1} z_{k}z_{k+1}.
\end{equation}
\begin{figure}[t]
{\includegraphics[width=0.35 \textwidth]{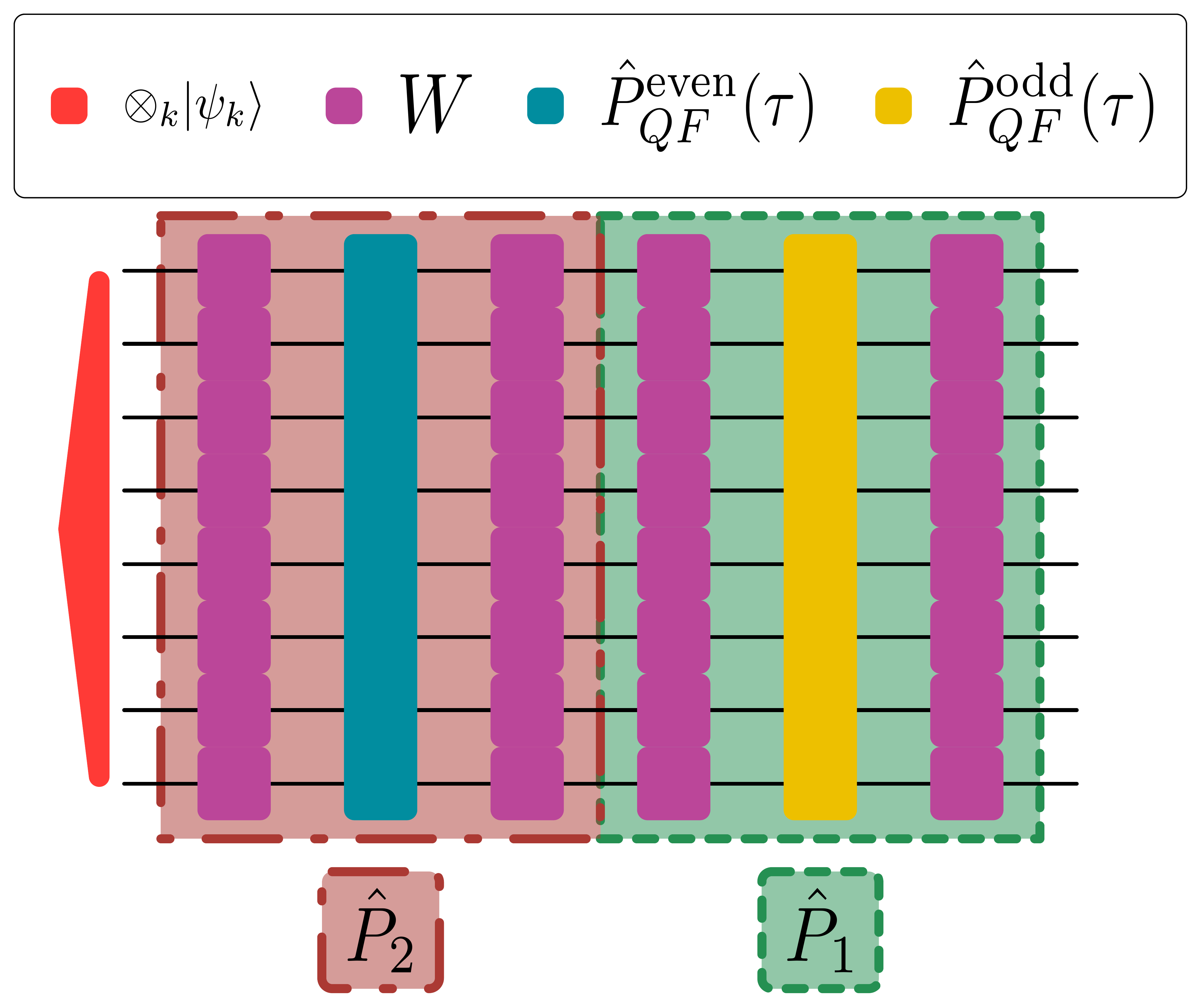}}
\caption{Digital-analog quantum circuit to simulate the evolution under Hamiltonian $H_{ZZ}$ for a time $\tau$. This simulation is carried out by transforming all qubits by $U_{QF}^{\text{even}}$, which entails transforming even qubits to the QF ($U_{QF}$) and odd qubits to the interaction picture ($U_{I}$), in a setup in which only even qubits are being driven. In this scenario, the analog propagator $\hat{P}_{QF}^{\text{even}}$ -- which describes the evolution under analog Hamiltonian $H_{QF}^{\text{even}}$ -- is then conjugated by Hadamard gates ($W$) on all qubits. This segment of the circuit is highlighted in dashed-dotted red line, and it simulates the evolution given by $\hat{P}_{2}(\tau)$. The circuit is repeated with the QF transformation being applied to odd qubits and the interaction picture transformation to even qubits ($U_{QF}^{\text{odd}}$), while only odd qubits are being driven. This segment, highlighted by a dotted green line, simulates the evolution given by $\hat{P}_{1}(\tau)$.}
\label{fig3}
\end{figure}
This sequence for simulating the evolution of $H_{ZZ}$ can be interpreted as the combination of two blocks: the first one represents the evolution given by $\hat{P}_{1}=e^{-iH_{1}t}$, where we only drive odd qubits, and the second one represents the evolution given by $\hat{P}_{2}=e^{-iH_{2}t}$, where we only drive even qubits, both in a frame rotated by Hadamard gates. The integrity of these simulation blocks relies on the fact that $[H_{1},H_{2}]=0$, meaning that the pairwise combination of propagators is exact. Then, the propagators corresponding to the two blocks can exactly describe the evolution of the whole, 
\begin{equation}
\hat{P}_{ZZ} = e^{-i H_{ZZ}t} = e^{-i (H_{1}+H_{2}) t} = \hat{P}_{1}\hat{P}_{2}.
\end{equation}
The propagator corresponding to $H_{ZZ}$ is computed as
\begin{equation}
\hat{P}_{ZZ}|\psi\rangle = \hat{P}_{1}\hat{P}_{2}|\psi\rangle = U^{\dagger}_{W} \hat{P}_{QF}^{\text{odd}} U_{W} U^{\dagger}_{W} \hat{P}_{QF}^{\text{even}} U_{W}|\psi\rangle
\end{equation}
where $\hat{P}_{QF}^{\text{odd}}$ and $\hat{P}_{QF}^{\text{even}}$ are the propagators generated by $H_{QF}^{\text{odd}}$ and $H_{QF}^{\text{even}}$, respectively. The former is achieved by rotating all qubits by $U_{\text{QF}}^{\text{odd}}$, and the latter is achieved by rotating all qubits by $U_{\text{QF}}^{\text{even}}$. Furthermore, $U_{W} = \bigotimes_{k} W_{k}$, where $W_{k} = e^{i\pi/2} e^{-i\pi y_{k}/4}e^{-i\pi z_{k}/2}$ represents the application of a Hadamard gate on qubit $k$. The simulation protocol is
\begin{enumerate}
\item Prepare an initial product state $\otimes_{k}|\psi_{k}\rangle$.
\item Apply Hadamard gates $W$ on all qubits.
\item Let the states evolve according to the underlying analog Hamiltonian with analog propagator $\hat{P}_{QF}^{\text{even}}(\tau)$ for time $\tau$.
\item Let the states evolve according to the underlying analog Hamiltonian with analog propagator $\hat{P}_{QF}^{\text{odd}}(\tau)$ for time $\tau$.
\item Apply Hadamard gates $W$ on all qubits.
\end{enumerate}
Note that operations that consist on the application of a unitary $U$, followed by their inverse $U^{\dagger}$, render the identity as the result, and thus are not mentioned in the simulation protocol steps. However, these operations are included in the figures for illustrative purposes. Due to the idiosyncrasies of the Hamiltonians derived in this protocol, we benefit from the absence of Trotter error, which implies no limits on application time, $\tau$, of the block. This block is represented in Fig.~\ref{fig3}. To evolve a state $|\psi\rangle$ with Hamiltonian $H_{ZZ}$, one must re-apply the block $M$ times where the total evolution time is $T=M\tau$. Gate-based quantum circuits describes the application of quantum gates following the usual flow of time. That is, from left to right, following the order in which the operators are applied on a quantum state represented by a ket.


\subsection{$XY$ model}

Let us now describe a protocol to simulate a $XY$ model in which all adjacent spins interact by $xx + yy$ terms. 

\subsubsection{1D Simulation}
\begin{figure}[t]
{\includegraphics[width=0.5 \textwidth]{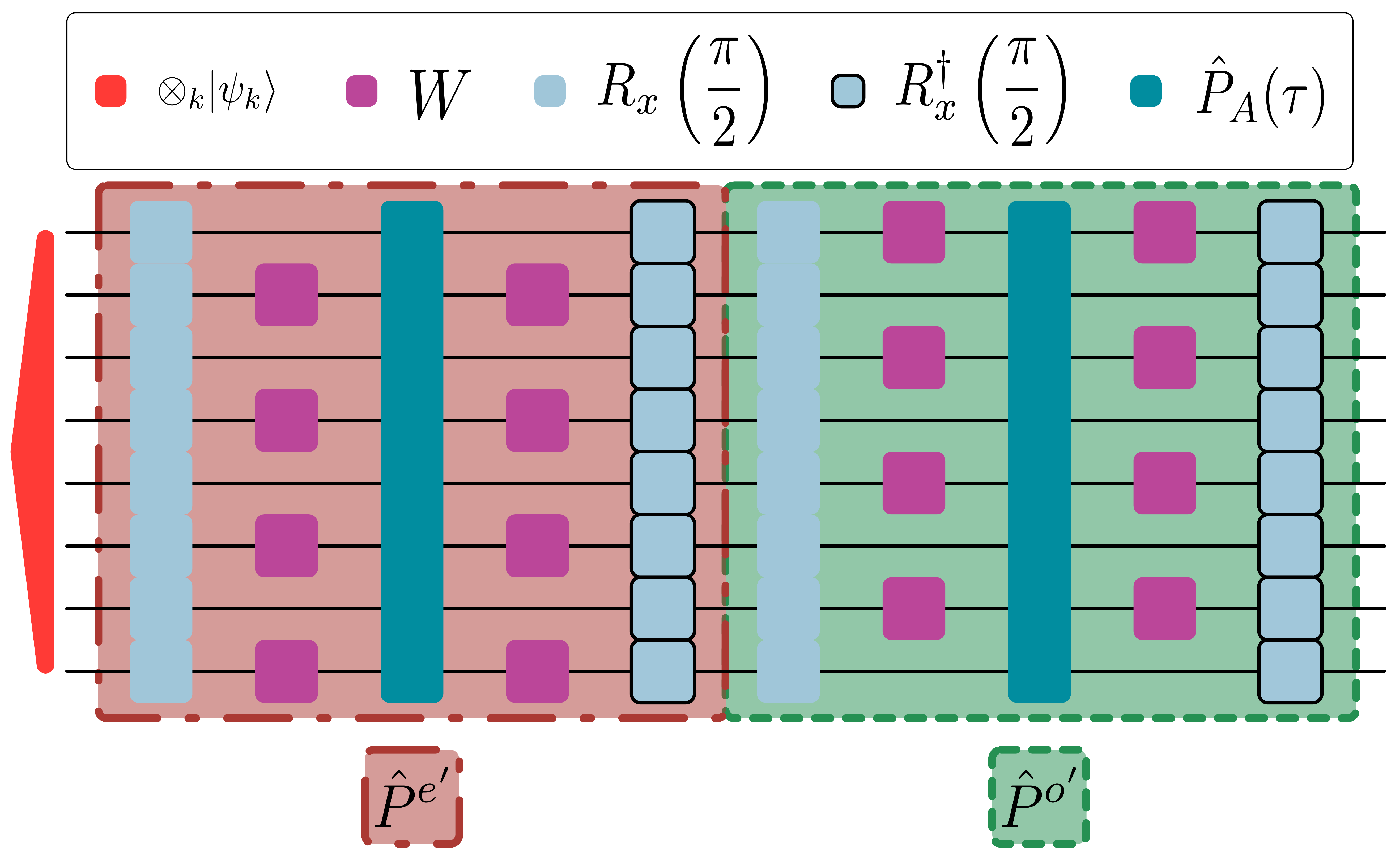}}
\caption{Digital-analog quantum circuit to simulate the evolution of an initial quantum state under Hamiltonian $H_{XY}$ for a time $\tau$. This simulation is carried out by conjugating the analog propagator $\hat{P}_{A}$ -- which describes the evolution under analog Hamiltonian $H_{A}$ -- by Hadamard gates ($W$) on even qubits, combined with $x$ rotations by $R_{x}(\pi/2)$ on all qubits. This segment of the circuit is highlighted in a dashed-dotted red line, and it simulates the evolution given by $\hat{P}^{e'}(\tau)$. The circuit is then repeated with Hadamard gates being applied to odd qubits, and this segment, highlighted in a dotted green line, simulates the evolution given by $\hat{P}^{o'}(\tau)$.}
\label{fig4}
\end{figure}

In the 1D case we start from the Hamiltonians in Eqs.~\ref{H_even},~\ref{H_odd}. By performing a global x-$\pi/2$ rotation, i.e. the same $R_{x}(\pi/2) \equiv R$ about each qubit, we find
\begin{eqnarray}
\nonumber H^{o'} &=& R^{\dagger} H^{o} R =  J\sum_{k=1}^{\frac{N}{2}}y_{2k-1}y_{2k} + J\sum_{k=1}^{\frac{N-1}{2}}x_{2k} x_{2k+1}, \\ 
H^{e'} &=& R^{\dagger} H^{e} R =  J\sum_{k=1}^{\frac{N}{2}}x_{2k-1}x_{2k} + J\sum_{k=1}^{\frac{N-1}{2}}y_{2k} y_{2k+1} 
\end{eqnarray}
which, upon summing, realize the 1D $XY$ chain Hamiltonian
\begin{equation}
H_{XY} = H^{e'}+ H^{o'}  = J\sum_{k=1}^{N-1}(x_{k}x_{k+1} + y_{k} y_{k+1}).
\end{equation}
The key to this protocol is that $[H^{e'},H^{o'}]=R^{\dagger}[H^{e},H^{o}]R=0$, which implies
\begin{equation}
\hat{P}_{XY} = e^{-i H_{XY}t} = e^{-i(H^{o'} + H^{e'})t} = \hat{P}^{o'}\hat{P}^{e'}.
\end{equation}
This allows us to decompose the total $XY$ propagator into the product of two toggled Hamiltonians which results in a Trotter-error-free dynamics simulation protocol. The propagator $\hat{P}_{XY}$ is further decomposed as
\begin{equation}
\hat{P}_{XY}|\psi\rangle = \hat{P}^{o'}\hat{P}^{e'}|\psi\rangle = U^{o'\dagger}\hat{P}_{A} U^{o'}U^{e'\dagger}\hat{P}_{A} U^{e'}|\psi\rangle,
\end{equation}
where $\hat{P}_{A}$ is the original analog propagator generated by $H_{A}$, of Eq.~\ref{H_control}, and $U^{o'} = \bigotimes_{k \,\text{odd}} W_{k}R_{k}R_{k+1}$, $U^{e'} = \bigotimes_{k \,\text{even}} R_{k-1}W_{k}R_{k}$. $W_{k}$ and $R_{k}$ represent the application of a Hadamard gate and a $\pi/2$ x-rotation, respectively, on qubit $k$. The simulation protocol is
\begin{enumerate}
\item Prepare an initial product state $\otimes_{k}|\psi_{k}\rangle$.
\item Apply a x-$\pi/2$ rotation on all qubits with $R_{x}(\pi/2)$. 
\item Apply Hadamard gates $W$ on even qubits.
\item Let the states evolve according to the underlying analog Hamiltonian with analog propagator $\hat{P}_{A}(\tau)$ for time $\tau$.
\item Apply Hadamard gates $W$ on all qubits.
\item Let the states evolve according to the underlying analog Hamiltonian with analog propagator $\hat{P}_{A}(\tau)$ for time $\tau$.
\item Apply Hadamard gates $W$ on odd qubits.
\item Undo the x-$\pi/2$ rotation on all qubits by $R_{x}^{\dagger}(\pi/2)$.
\end{enumerate}
The entire sequence of operations needed to evolve by the $XY$ Hamiltonian is depicted in Fig.~\ref{fig4}. To evolve for a total time $T$ with Hamiltonian $H_{XY}$, one must re-apply the block $M=T/\tau$ times. Note that the three layers of single qubit rotations in between evolution by the analog propagators simplify into the product of single qubit gates, which in this case simplifies to $R^{\dagger}W R = (x+y)/\sqrt{2}$.

\subsubsection{2D Simulation and Digital vs. Digital-Analog Trotter Errors}
\label{sec:2D_XY}

The same two-Hamiltonian decomposition may be performed in two-dimensions, taking the Hamiltonians in Eqs.~\ref{eq:H_{I}},~\ref{eq:H_{II}}, such that $H^{2D}_{XY}=H_{I}+H_{II}$. However, since $[H_{I},H_{II}] \ne 0$ in two dimensions, we must resort to an approximate Trotter decomposition of the 2D $XY$ propagator. If we compute $[H_{I},H_{II}]$ we find 16 non-commuting terms, as shown in Table~\ref{table1}.

\begin{table*}[]
\centering
\begin{tabular}{| >{\centering\arraybackslash} m{1.8 cm}| |*{8}{>{\centering\arraybackslash} m{1.8 cm}|}|@{}m{0pt}@{}}\hline
\diagbox{$H_{I}$}{$H_{II}$} & \makecell{$x_{2i'-1,2j'}$ \\ $x_{2i'-1,2j'+1}$} & \makecell{$x_{2i'-1,2j'}$ \\ $x_{2i',2j'}$} & \makecell{$x_{2i',2j'-1}$ \\ $x_{2i',2j'}$} & \makecell{$x_{2i',2j'-1}$ \\ $x_{2i'+1,2j'-1}$} & \makecell{$y_{2i'-1,2j'-1}$ \\ $y_{2i'-1,2j'}$} & \makecell{$y_{2i'-1,2j'-1}$ \\ $y_{2i',2j'-1}$} & \makecell{$y_{2i',2j'}$ \\ $y_{2i',2j'+1}$} & \makecell{$y_{2i',2j'}$ \\ $y_{2i'+1,2j'}$} &\\ [10pt] \hline\hline
\makecell{$x_{2i-1,2j-1}$ \\ $x_{2i-1,2j}$} & 0 & 0 & 0 & 0 & 0 & $T_{25}(x\leftrightarrow y)$ & 0 & $T_{45}(x\leftrightarrow y)$ &\\ [10pt] \hline
\makecell{$x_{2i-1,2j-1}$ \\ $x_{2i,2j-1}$} & 0 & 0 & 0 & 0 & \makecell{$x_{2i,2j-1}$ \\ $y_{2i-1,2j}$ \\ $z_{2i-1,2j-1}$} & 0 & $T_{36}(x\leftrightarrow y)$ & 0 &\\ [10pt] \hline
\makecell{$x_{2i,2j}$ \\ $x_{2i,2j+1}$} & 0 & 0 & 0 & 0 & 0 & \makecell{$x_{2i,2j}$ \\ $y_{2i-1,2j+1}$ \\ $z_{2i,2j+1}$} & 0 & $T_{47}(x\leftrightarrow y)$ &\\ [10pt] \hline
\makecell{$x_{2i,2j}$ \\ $x_{2i+1,2j}$} & 0 & 0 & 0 & 0 & \makecell{$x_{2i,2j}$ \\ $y_{2i+1,2j-1}$ \\ $z_{2i+1,2j}$} & 0 & \makecell{$x_{2i+1,2j}$ \\ $y_{2i,2j+1}$ \\ $z_{2i,2j}$} & 0 &\\ [10pt] \hline
\makecell{$y_{2i-1,2j}$ \\ $y_{2i-1,2j+1}$} & 0 & \makecell{$- x_{2i,2j}$ \\ $y_{2i-1,2j+1}$ \\ $z_{2i-1,2j}$} & 0 & \makecell{$- x_{2i,2j+1}$ \\ $y_{2i+1,2j}$ \\ $z_{2i+1,2j+1}$} & 0 & 0 & 0 & 0 &\\ [10pt] \hline
\makecell{$y_{2i-1,2j}$ \\ $y_{2i,2j}$} & $T_{52}(x\leftrightarrow y)$ & 0 & \makecell{$- x_{2i,2j-1}$ \\ $y_{2i-1,2j}$ \\ $z_{2i,2j}$}  & 0 & 0 & 0 & 0 & 0 &\\ [10pt] \hline
\makecell{$y_{2i,2j-1}$ \\ $y_{2i,2j}$} & 0 & $T_{63}(x\leftrightarrow y)$ & 0 & \makecell{$- x_{2i+1,2j-1}$ \\ $y_{2i,2j}$ \\ $z_{2i,2j-1}$} & 0 & 0 & 0 & 0 &\\ [10pt] \hline
\makecell{$y_{2i,2j-1}$ \\ $y_{2i+1,2j-1}$} & $T_{54}(x\leftrightarrow y)$ & 0 & $T_{74}(x\leftrightarrow y)$ & 0 & 0 & 0 & 0 & 0 &\\  [10pt] \hline
\end{tabular}
\caption{Table containing the commutators between the different toggled two-body interactions described in $H_{I}$ and $H_{II}$, up to a global factor of $2i$. The extension of the toggled Hamiltonians from a 1-dimensional chain to a 2-dimensional lattice implies that there will be some non-commuting terms, as is reflected in this table. The objects $T_{ij}$ refer to the table elements from the $i$th row and $j$th column. Note that Eq.~\ref{eq:A} (\ref{eq:B}) correspond to a summation over the elements of the bottom left (top right) blocks of this table.}
\label{table1}
\end{table*}

Let us now compare the errors arising from a first order Trotter decomposition of our target evolution unitary. 
Overall, our goal is to determine the gate complexity of an approximate product decomposition $U_{PD}$ such that $||U_T(\tau) - U_{PD}(\tau)|| \le \epsilon$ for an $\epsilon$ of our choosing. Here the target propagator is generated by exponentiating the target Hamiltonian $H^{2D}_{XY}$ while $U_{PD}$ is generated by a first order Trotter decomposition which may be implemented through our DA Hamiltonians or through a digitized decomposition.

A first order Trotterization approximates an operator exponential of two generally non-commuting operators, $\alpha$ and $\beta$, as  $e^{\Delta t \alpha} e^{\Delta t \beta} = e^{\Delta t (\alpha+\beta)} + \mathcal{O}(\Delta t^2 [\alpha,\beta])$ by discarding the $\Delta t^2$ terms in the small $\Delta t$ regime. This quantity can be made arbitrarily small by breaking up the total evolution time into sufficiently small pieces $\Delta t=\tau/N$. Bounding the error in the DA case reduces to computing $||[H_{I},H_{II}]||$. Breaking down each Hamiltonian into its $X$ and $Y$ components, such that $H_{i} = H^{XX}_{i} + H^{YY}_{i}$, simplifies the commutator norm to $||[H^{YY}_{I},H^{XX}_{II}] + [H^{XX}_{I},H^{YY}_{II}]||=||A+B||$. See that $A$ is composed by the terms given in the $4\times 4$ grid in the bottom left of Table~\ref{table1}, whereas B is composed by those terms in the top right $4\times 4$ grid. These operators are 

\begin{figure}[t]
    \centering
    \includegraphics[width=0.9 \columnwidth]{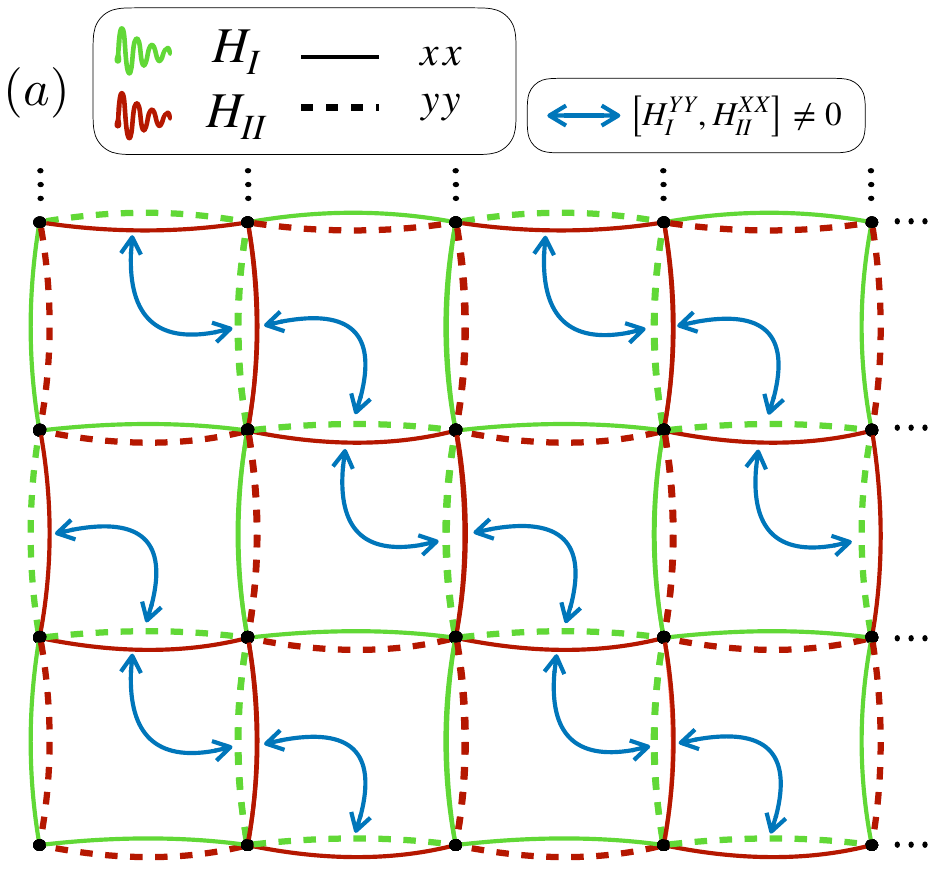}
    \includegraphics[width=0.9 \columnwidth]{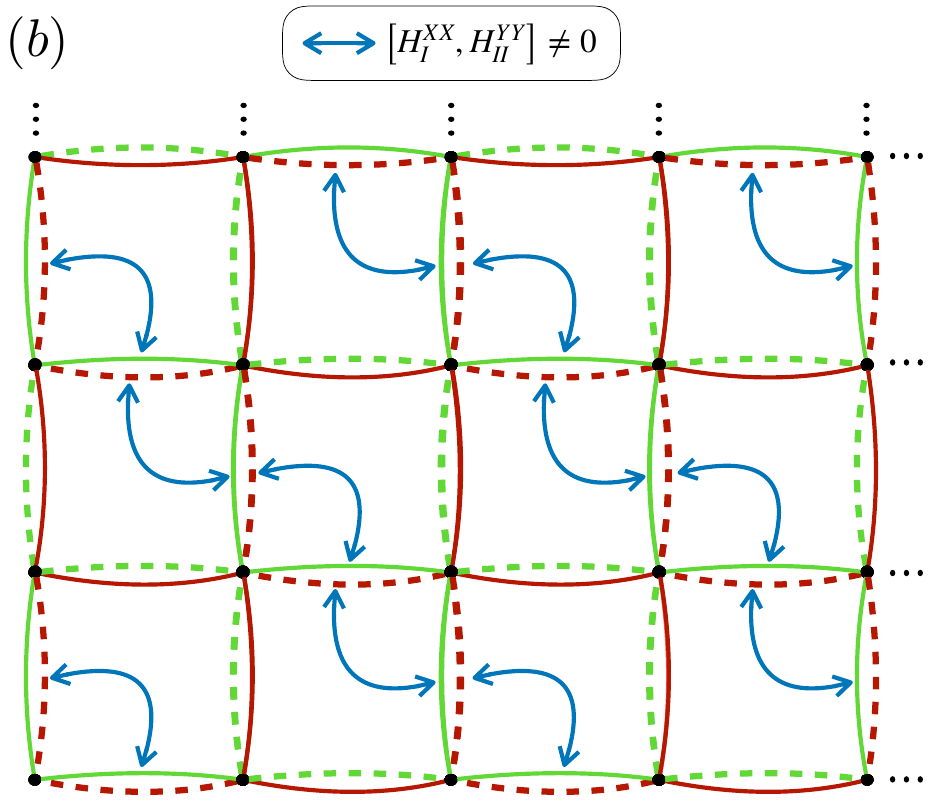}
    \caption{Lattice representation of the interactions featured on $H_{I}$ (green) and $H_{II}$ (red), where the blue arrows indicate the non-commuting terms between $H_{I}$ and $H_{II}$. These Hamiltonians are split into $xx$ and $yy$ - supported operators, $H_{i} = H_{i}^{XX} + H_{i}^{YY}$, in order to identify the two non-commuting operators: (a) Non-zero terms of $[H_{I}^{YY},H_{II}^{XX}]$, (b) Non-zero terms of $[H_{I}^{XX},H_{II}^{YY}]$. Jointly, these terms estimate the total Trotter error of the DA decomposition.}
    \label{fig5}
\end{figure}

\begin{widetext}
\begin{eqnarray}
\label{eq:A}
A &=& J^{2} \sum_{i,j}\sum_{i',j'} [y_{2i,2j-1} y_{2i+1,2j-1} + y_{2i-1,2j} y_{2i,2j} + y_{2i,2j-1} y_{2i,2j} + y_{2i-1,2j} y_{2i-1,2j+1}, \nonumber \\
  &&  x_{2i',2j'-1} x_{2i'+1,2j'-1}+x_{2i'-1,2j'} x_{2i',2j'} + x_{2i',2j'-1} x_{2i',2j'} + x_{2i'-1,2j'} x_{2i'-1,2j'+1}] \nonumber \\ 
  &=& - 2i J^{2} \sum_{i,j} [x_{2i,2j+1}y_{2i+1,2j}z_{2i+1,2j+1} + x_{2i+1,2j-1}y_{2i,2j}z_{2i,2j-1} + x_{2i,2j}y_{2i-1,2j+1}z_{2i-1,2j} \nonumber \\
  &+& x_{2i,2j-1}y_{2i-1,2j}z_{2i,2j} + (y \leftrightarrow x)], 
\end{eqnarray}
\begin{eqnarray}
\label{eq:B}
\nonumber B &=& J^{2} \sum_{i,j} \sum_{i',j'} [x_{2i-1,2j-1} x_{2i-1,2j} + x_{2i-1,2j-1} x_{2i,2j-1} + x_{2i,2j} x_{2i,2j+1} + x_{2i,2j} x_{2i+1,2j}, \nonumber \\
&&  y_{2i'-1,2j'-1} y_{2i'-1,2j'} + y_{2i'-1,2j'-1} y_{2i',2j'-1} + y_{2i',2j'} y_{2i',2j'+1} + y_{2i',2j'} y_{2i'+1,2j'}] \nonumber \\ 
\nonumber &=& 2i J^{2} \sum_{i,j} [x_{2i,2j}y_{2i-1,2j+1}z_{2i,2j+1} + x_{2i+1,2j}y_{2i,2j+1}z_{2i,2j} + x_{2i,2j}y_{2i+1,2j-1}z_{2i+1,2j} \\
&+& x_{2i,2j-1}y_{2i-1,2j}z_{2i-1,2j-1} + (y \leftrightarrow x)].
\end{eqnarray}

\end{widetext}
Alternatively, from visually inspecting supports and Pauli character of the Hamiltonians $H_I$ and $H_{II}$ denoted in red and green in Fig.~\ref{fig5}, we can see that there are 8 terms per unit cell in A and that there are likewise 8 similar, but differently supported terms in B. Summing over the two sets of terms in the bulk, we obtain 
\begin{widetext}

\begin{eqnarray}
    \nonumber ||[H_{I},H_{II}]|| &=& ||2i J^{2} \sum_{i,j} (-1)^{i+j}  z_{i,j} [(x_{i-1,j} y_{i,j-1} - x_{i,j+1} y_{i+1,j}) + (x\leftrightarrow y)] || \\ 
    \nonumber &\le& 2 J^{2} \sum_{i,j} ||(-1)^{i+j} z_{i,j} [(x_{i-1,j} y_{i,j-1} - x_{i,j+1} y_{i+1,j}) + (x\leftrightarrow y)] || \\
    &=& 2 J^{2} N^2 ||z_{i,j} [(x_{i-1,j} y_{i,j-1} - x_{i,j+1} y_{i+1,j}) + (x\leftrightarrow y)]|| \leq 8 J^{2} N^2,
    \label{eq:DATrottercommutator}
\end{eqnarray}
\end{widetext}

where we have used the triangle inequality on the spectral norms of the operators. 

In order to get a better insight on the performance of the DA computation of the two-dimensional $XY$ model, we need to compare the Trotter error of both digital and DA approaches. This error is proportional to the commutator of $\left[H _I,H_{II} \right]$ given in Eq.~\ref{eq:DATrottercommutator} in the DA case. In the purely digital case, the commutator we need to compute is $\left[H_{xx},H_{yy} \right]$, where $H_{xx}$ contains all $xx$ qubit interactions and $H_{yy}$ all the $yy$ interactions. Independent of the order in which the gates are implemented, the digital error is bounded by 
\begin{eqnarray}
\label{eq:DD_comm}
\nonumber ||[H_{xx},H_{yy}]|| &=& || J^{2} \sum_{i,j} [(x_{i,j} x_{i+1,j} + x_{i,j} x_{i,j+1}) ,  (x\leftrightarrow y)] || \\ 
&\leq& 24 J^{2} N^2
\end{eqnarray}
where the final factor arises from a product of the factor of two for the $N^2$ vertical and horizontal edges, a factor of $6$ counting all the non-commuting $yy$ neighbors of each $xx$ interaction, and a final factor of two arising from the Pauli commutation relations. Alternatively, by analyzing the forms of Eqs.~\ref{eq:A} and \ref{eq:B} we note that the $A$ and $B$ components of the commutator can be identified with free Fermions hopping along the diagonal loops of the two-dimensional lattice as defined by the blue arrows in Fig.~\ref{fig5}. Next we Jordan-Wigner transform to a majorana representation, take periodic boundary conditions, and Fourier transform along the loops. As a result, the spectral norm of $A$ and $B$ is tightened from $\mathcal{O}(NJ)^2 \rightarrow \mathcal{O}(J)^2$ which removes the extensive factor. This tighter bound is proved in Appendix~\ref{app:norm}. Likewise we may use similar techniques to decompose the Digital commutator of Eq.~\ref{eq:DD_comm} into a sum of three times as many free fermion Hamiltonians. The resulting ratio of purely digital to digital-analog commutator norms is still a factor of three. In either case, the DA protocol improves the Trotter error bound by a constant factor of three. This constant factor speedup can be used to extend the simulation time by the same factor.


\subsection{Heisenberg model}
\begin{figure}[t]
{\includegraphics[width=0.5 \textwidth]{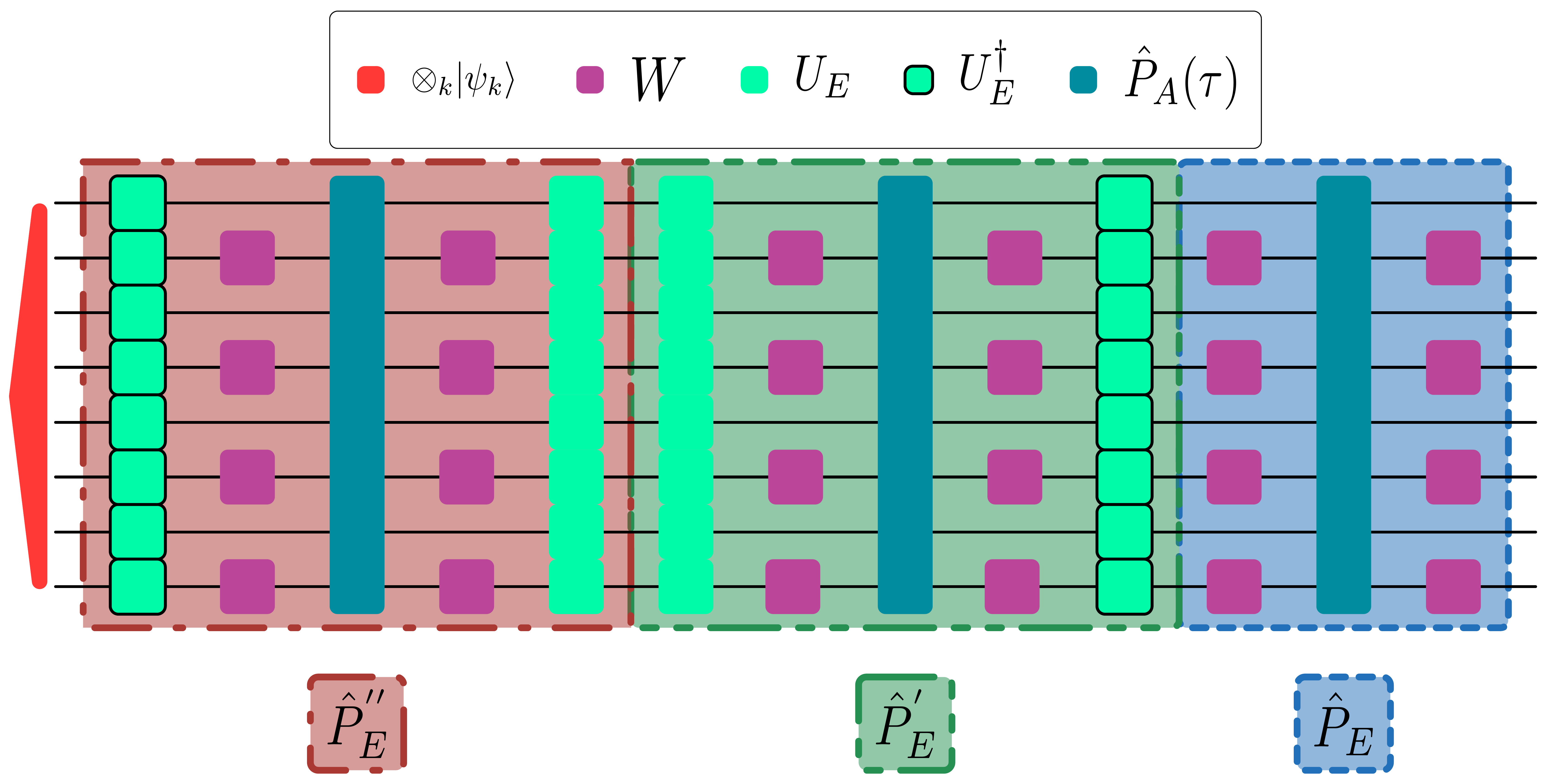}}
\caption{Digital-analog quantum circuit to simulate the evolution of an initial quantum state under Hamiltonian $H_{\text{Heis}}$ for a time $\tau$. This simulation is carried out by conjugating the analog propagator $\hat{P}_{A}$ -- which describes the evolution under analog Hamiltonian $H_{A}$ -- by Hadamard gates ($W$) on even qubits. It is additionally conjugated by $U_{E}^{2} = -U_{E}^{\dagger}$, where $U_{E}$ is a cyclic transformation that allows us to obtain all $S_{i}S_{i+1}$ interactions. This segment of the circuit is highlighted in red, and it simulates the evolution given by $\hat{P}_{E}^{''}(\tau)$. The first repetition of the circuit is highlighted in a dashed-double dotted green line and it simulates the evolution given by $\hat{P}_{E}^{'}(\tau)$. This is done by conjugating $P_{A}(\tau)$ by Hadamard gates on even qubits, followed by a permutation by $U_{E}$. Then, the circuit is repeated a final time, solely conjugating the analog propagator by Hadamard gates on even qubits, to simulate the evolution given by $\hat{P}_{E}(\tau)$. This last segment is highlighted in a dotted blue line.}
\label{fig6}
\end{figure}

We now consider the task of simulating the more complex Heisenberg spin model. The Hamiltonian describing the Heisenberg chain in 1 dimension is $H_{\text{Heis}}= \sum_i \bm{S}_i \cdot \bm{S}_{i+1}$, with $\bm{S}=(x,y,z)$. Consider the Bloch sphere rotation $U_{E}=e^{-i\theta(x+y+z)}$. We can set the angle $\theta$ such that this rotation becomes cyclic; that is, $\theta=\pi/3\sqrt{3}$ leads to a cyclic permutation $x\rightarrow z$, $y\rightarrow x$ and $z\rightarrow y$. This transformation is realized by 
\begin{equation}
U_{E}=e^{-i\frac{\pi}{3\sqrt{3}}(x+y+z)} = \frac{1}{2}[\mathbb{1}-i(x+y+z)]
\end{equation}
which can easily be implemented on individual qubits by the Euler decomposition $U_E=e^{-i y \frac{ \pi }{4}} e^{-i z \frac{ \pi }{4}}$. The cyclic nature of this transformation is manifested through the property $U_{E}^{3}=-\mathbb{1}$. Applying this transformation on all qubits on the Hamiltonian in Eq.~\ref{H_even} zero, one, and two times, leads to the following Hamiltonians,
\begin{eqnarray}
H_E &=& H^{e} = J\sum_{k=1}^{\frac{N}{2}}x_{2k-1}x_{2k} + J\sum_{k=1}^{\frac{N-1}{2}}z_{2k} z_{2k+1}, \\
\nonumber H_E' &=& U_{E}^{\dagger}H^{e}U_{E} = J\sum_{k=1}^{\frac{N}{2}}z_{2k-1}z_{2k} + J\sum_{k=1}^{\frac{N-1}{2}}y_{2k} y_{2k+1}\\
\nonumber H_E''&=& U_{E}^{2 \dagger}H^{e}U_{E}^{2} = J\sum_{k=1}^{\frac{N}{2}}y_{2k-1}y_{2k} + J\sum_{k=1}^{\frac{N-1}{2}}x_{2k} x_{2k+1}.
\end{eqnarray}
Summing them together, we obtain the Heisenberg Hamiltonian,
\begin{equation}
H_{\text{Heis}} = H_{E} + H_{E}'+H_{E}'' = J\sum_{k=1}^{N-1}(x_{k}x_{k+1}+y_{k}y_{k+1}+z_{k}z_{k+1}).  
\end{equation}
In this case, the Hamiltonians do not commute with each other, which means that the construction of the propagator will include Trotter error (analyzed below),
\begin{eqnarray}
\nonumber \hat{P}_{\text{Heis}} &=& e^{-i H_{\text{Heis}}t} = e^{-i(H_{E} + H'_{E} + H''_{E})t} \\
&=& \hat{P}_{E}\hat{P}'_{E}\hat{P}''_{E} + \mathcal{O}(J^2 t^{2}).
\end{eqnarray}
The propagator $\hat{P}_{\text{Heis}}$ is constructed as
\begin{eqnarray}
\hat{P}_{\text{Heis}}|\psi\rangle &\approx& \hat{P}_{E}\hat{P}'_{E}\hat{P}''_{E}|\psi\rangle \\
\nonumber &=& U^{e \dagger}\hat{P}_{A} U^{e} U_{E}^{\dagger}U^{e \dagger}\hat{P}_{A} U^{e}U_{E} U_{E}^{2 \dagger}U^{e \dagger}\hat{P}_{A} U^{e}U_{E}^{2} |\psi\rangle,
\end{eqnarray}
where $\hat{P}_{A}$ is the analog propagator generated by $H_{A}$, and $U^{e} = \bigotimes_{k \,\text{even}} \mathbb{1}_{k-1}W_{k} = U^{e \dagger}$. $W_{k}$ represents the application of a Hadamard gate on qubit $k$. This protocol is
\begin{enumerate}
\item Prepare an initial product state $\otimes_{k}|\psi_{k}\rangle$.
\item Apply the cyclic transformation twice with $U_{E}^{2}$, which is equivalent to $U_{E}^{\dagger}$, on all qubits.
\item Apply Hadamard gates $W$ on even qubits.
\item Let the states evolve according to the underlying analog Hamiltonian with analog propagator $\hat{P}_{A}(\tau)$ for time $\tau$.
\item Apply Hadamard gates $W$ on even qubits.
\item Undo the double cyclic transformation by applying $U_{E}$ on all qubits.
\item Apply the cyclic transformation with $U_{E}$ on all qubits.
\item Apply Hadamard gates $W$ on even qubits.
\item Let the states evolve according to the underlying analog Hamiltonian with analog propagator $\hat{P}_{A}(\tau)$ for time $\tau$.
\item Apply Hadamard gates $W$ on even qubits.
\item Undo the cyclic transformation with $U_{E}^{\dagger}$ on all qubits.
\item Apply Hadamard gates $W$ on even qubits.
\item Let the states evolve according to the underlying analog Hamiltonian with analog propagator $\hat{P}_{A}(\tau)$ for time $\tau$.
\item Apply Hadamard gates $W$ on even qubits.
\end{enumerate}
This sequence of quantum gates constitutes a block, which can be seen in Fig.~\ref{fig6}. To evolve with Hamiltonian $H_{\text{Heis}}$ for a total time $T$, one must re-apply the block $M=T/\tau$ times. 

\subsubsection{Digital vs. Digital-Analog Synthesis Errors}

In order to quantify the computational benefit of this method, let us compute and compare the above Trotterized error against that of a digitized two-local decomposition. A digitized decomposition we will employ alternating layers of $xx$, $yy$, and $zz$ interactions applied to all even bonds, followed by the same operator action on odd bonds. Such a decomposition is based on the fact that all interactions, on a single bond, commute, but the interactions on adjacent bonds, which share a single spin, do not commute. To first order, the Trotter error is given as 
\begin{eqnarray}
J^2\norm{[\bm{S}_{i-1}\cdot \bm{S}_i,\bm{S}_i \cdot \bm{S}_{i+1}]} &=& J^2\norm{\sum_{\mu,\nu}\sigma^\mu_{i-1} [\sigma^\mu_i,\sigma^\nu_i] \sigma^\nu_{i+1}} \nonumber \\
&=& 2 J^2\norm{\bm{S}_{i-1}\cdot\bm{S}_i \cross \bm{S}_{i+1}} \nonumber \\
 & \le & 12J^2. 
\end{eqnarray}
where we have used the fact that $\bm{S}_{i-1}\cdot\bm{S}_i \cross \bm{S}_{i+1}$ contains 6 Pauli terms. For a 1D Heisenberg chain the total commutator is bounded by $12 J^2 N$. 
Meanwhile on the DA side we need to bound 
\begin{eqnarray}
e^{-it (H_E+H_E'+H_E'')} &=& e^{-it H_E} e^{-it (H_E'+H_E'')} \\ &+& \mathcal{O}(t^2 [H_E,H_E'+H_E''])\nonumber \\ 
&=& e^{-it H_E} e^{-it H_E'} e^{-it H_E''}\nonumber  \\ 
&+& \mathcal{O}(t^2 ([H_E,H_E'+H_E''] + [H_E',H_E''])) \nonumber 
\end{eqnarray}
These commutators are
\begin{eqnarray}
[H_{E},H'_{E}] &=& 2 i J^2\sum_{k} x_{k-1}z_{k}y_{k+1}, \\
\nonumber [H_{E},H''_{E}] &=& -2 i J^2\sum_{k} z_{k-1}x_{k}y_{k+1}, \\
\nonumber [H'_{E},H''_{E}] &=& 2 i J^2\sum_{k} z_{k-1}y_{k}x_{k+1},
\end{eqnarray}
and their sum can be bounded by
\begin{eqnarray}
\nonumber && ||[H_{E},H'_{E}] + [H_{E},H''_{E}] + [H'_{E},H''_{E}]|| \\
\nonumber &=& 2 J^{2} ||\sum_{k} x_{k-1}z_{k}y_{k+1} - z_{k-1}x_{k}y_{k+1} + z_{k-1}y_{k}x_{k+1}|| \\
\nonumber & \leq & 2 J^{2} \sum_{k} || x_{k-1}z_{k}y_{k+1} - z_{k-1}x_{k}y_{k+1} + z_{k-1}y_{k}x_{k+1}|| \\
&\leq& 6 J^{2}N.
\end{eqnarray}
We again find that the bound on the error in the DA protocol is smaller by a constant factor than in the digital approach.

\section{Practical Implementation}
In order to experimentally realize our DA simulation protocols in an accurate manner further practical experimental steps are required. The critical steps for doing so, whose details depend on the user's specific goals, are broadly partitioned as either i) characterization or ii) Hamiltonian optimization. Since each of these steps bring their own theoretical and experimental challenges, we now describe promising paths forward for each step. 

A critical step towards validating the accuracy of DA simulations, thereby quantifying their error, is to accurately characterize the analog many-body Hamiltonian at the center of our protocols. While conceptually simple, the characterization of a many-body Hamiltonian is not scalable (with exponentially growing complexity) by naive process tomography~\cite{Nielsen2000}. To aid in the scalable characterization of our Hamiltonians, we note that all of the expected interactions are geometrically local and, using this information, one should take advantage of Hamiltonian tomography schemes with polynomial growing model spaces as constricted by locality
~\cite{Shabani2011, daSilva2011, Qi2017, Bairey2019}. Hamiltonian estimation is further complicated by interactions coupling the principle system to unwanted environmental degrees of freedom and, to address this complication, we point the interested reader to recently developed open quantum systems characterization techniques
~\cite{Bairey2020, Dumitrescu2020}. Additionally, Bayesian Hamiltonian learning
~\cite{Granade2012} techniques may also be considered, although efficient importance sampling is required to adequately update models in this case. 

After experimentally identifying the dominant interactions, a natural next step is to eliminate unwanted couplings. Our analog Hamiltonian arises from a model relying on a two-level approximation and perturbation theory in $\Omega/\delta$. However, it is known that the CR-operation comes with a variety of additional terms~\cite{Sheldon2016,Magesan2020,Malekakhlagh2020}, such as $z_2, y_2, z_1 z_2$ as well as spectator phase errors, in practice. One may consider a few routes in order to combat these additional terms. For example, tailoring echo sequences can eliminate certain unwanted interactions~\cite{Ku2020} and, in addition, it has been shown that residual single qubit interactions can be removed by applying active cancellation tones
~\cite{Sheldon2016}. Another promising avenue for removing residual interactions comes from judiciously arranging, or actively controlling, qubit frequencies or their relative anharmonicities. For example, Ref.~\citenum{Malekakhlagh2020} provides a detailed analysis of the role qubit frequencies play and have shown that certain bands in the space of frequency detunings (see regions I and IV in Fig.~4 (f)) maximize the signal to noise ratio $\abs{\frac{zx}{zz}}$. Even more recent work~\cite{Kandala2020} has highlighted how additional fixed frequency coupling elements, which dress the qubit level spacings, may also remove unwanted $zz$ interactions. 

Lastly, instead of removing the residual couplings, one may leverage the additional interactions to define new classes of analog Hamiltonians. These analog Hamiltonians would be useful in simulating the dynamics of different spin models. In the limit that these additional terms are sufficiently small, one would expect them to contribute as disorder or small fluctuations in the system parameters. In this case, the (low energy theory and effective) model is expected to still lie in the parent model's universality class. Alternatively, outside this limit the presence of the additional terms may potentially enrich the computational capability of the analog Hamiltonian as applied to more complex spin models. 




\section{Discussions}


In this work, we start from a Hamiltonian based on the Rabi model describing two superconducting qubits interacting through the cross-resonance effect, and propose an extension to a multi-qubit scenario. The resulting Hamiltonian is transformed to a reference frame where only two-body interactions remain, resulting in our analog Hamiltonian. With it, we have assembled a Hamiltonian toolbox through toggling by different single-qubit gates. 

The variety of Hamiltonians we have obtained were efficiently combined to simulate Ising, $XY$, and Heisenberg spin models on a 1-dimensional chain, as well as the $XY$ model on a 2-dimensional lattice. For the 1D Ising and $XY$ models, our simulation protocols are Trotter-error free up to first order in $\Omega/\delta$, meaning that the full time evolution is given by a single DA block. For the 2D $XY$ and 1D Heisenberg chain, we were able to reduce the error in a first order Trotter approximation by a constant factor of $3$ for 2D $XY$ and of $2$ for the Heisenberg chain. Our techniques therefore extend the duration of possible time evolutions by a constant factor. While the constant factor improvement does not provide a polynomial speedup in the asymptotic limit, it does provide a meaningful and practical advantage for near term, noisy, simulations. A natural avenue of future research could be to explore the possible reach of quantum computation by offering a larger collection of analog Hamiltonians which naturally arise in superconducting platforms. Going beyond our simple Trotter analysis, it would also be interesting to investigate the scaling improvements resulting from the use of the DA Hamiltonians within more advanced product formulas \cite{Tran2020} or alternative Hamiltonian simulation techniques \cite{Berry2015}. 

Finally we have provided a succinct discussion regarding the steps which are necessary to implement our DA protocols in practice. In doing so we outlined promising routes towards scalable characterization and tailoring the precise nature of the analog interactions. Another issue which must be tackled is the problem of geometrically designing the qubit detunings such that all qubits are kept within a particular range. Then, given these detunings, one should increase or decrease the individual driving to maintain a constant ratio $\Omega/\delta$ for all neighboring pairs. In reality, one must go beyond this simple approximation and will need to calibrate each of the individual drivings as the cross-resonance interaction may be highly sensitive to resonances which depend not only on the detuning but also on the qubit's anharmonicities~\cite{Malekakhlagh2020}.

\acknowledgements 
The authors are grateful to Moein Malekakhlagh for helpful discussions regarding the cross-resonance gate. 
TG-R, RA-P, AM, LCC and MS acknowledge support from Spanish Government PGC2018-095113-B-I00 (MCIU/AEI/FEDER, UE) and Basque Government IT986-16, together with the projects QMiCS (820505) and OpenSuperQ (820363) of the EU Flagship on Quantum Technologies, as well as the EU FET Open Projects Quromorphic (828826) and Epiqus (899368). They also acknowledge support from the U.S. Department of Energy, Office of Science, Office of Advanced Scientific Computing Research (ASCR) quantum algorithm teams program, under field work proposal number ERKJ333. LCC would like to acknowledge the financial support from the Brazilian ministries MEC and MCTIC, funding agency CNPq, and the Brazilian National Institute of Science and Technology of Quantum Information (INCT-IQ). This study was financed in part by the Coordena\c{c}\~{a}o de Aperfei\c{c}oamento de Pessoal de N\'{i}vel Superior -- Brasil (CAPES) -- Finance Code 001. P.L. and E.F.D. acknowledges DOE ASCR funding under the Quantum Computing Application Teams program, FWP No. ERKJ347. A portion of this work was performed at Oak Ridge National Laboratory, managed by UT-Battelle, LLC, for the US Department of Energy under contract no. DE-AC05-00OR22725.

\appendix
\begin{widetext}

\section{CR Hamiltonian}
\label{app:CRHamiltonians}

In this appendix we provide the details of the derivation of the effective Hamiltonians described in Sec.~\ref{sec:CRHamiltonians}.

\subsection{Two qubit case}
\label{sec:2_qubit_details}
The transformation that takes the Hamiltonian in Eq.~\ref{eq:Ham2Q} into a doubly rotating frame is given by
\begin{equation}
U_{12} = \exp\left[ -\frac{i}{2}((\omega_{1}t + \phi_{1})z_{1} + (\omega_{2}t + \phi_{2})z_{2})\right].
\end{equation}
This operation results in
\begin{eqnarray}
\nonumber H_{2} &=& \frac{1}{2}(\delta_{1}z_{1} + \delta_{2}z_{2}) + \Omega_{1}\cos(\omega_{1}t+\phi_{1})[x_{1}\cos(\omega_{1}t+\phi_{1})-y_{1}\sin(\omega_{1}t+\phi_{1})] \\ 
\nonumber &+& \Omega_{2}\cos(\omega_{2}t+\phi_{2})[x_{2}\cos(\omega_{2}t+\phi_{2})-y_{2}\sin(\omega_{2}t+\phi_{2})] \\
&+& \frac{g}{2}[x_{1}\cos(\omega_{1}t+\phi_{1})-y_{1}\sin(\omega_{1}t+\phi_{1})][x_{2}\cos(\omega_{2}t+\phi_{2})-y_{2}\sin(\omega_{2}t+\phi_{2})],
\end{eqnarray}
with $\delta_{k} = \omega_{k}^{q} - \omega_{k}$. Next a rotating wave approximation (RWA) is performed by dropping terms proportional to $e^{\pm 2i\omega_{1}}$, $e^{\pm 2i\omega_{2}}$, and $e^{\pm i(\omega_{1}+\omega_{2})}$. The validity of this approximation relies on a time-average of the Hamiltonian and noting that $\Omega/(\omega_i+\omega_j)\ll 1$ and $g/(\omega_i+\omega_j)\ll 1$, $\forall i,j$. The remaining terms are either static, or rotating at $\pm(\omega_{1}-\omega_{2})$:
\begin{equation}
H_{2} = \frac{1}{2}(\delta_{1}z_{1} + \delta_{2}z_{2}) + \frac{1}{2}(\Omega_{1}x_{1} + \Omega_{2}x_{2}) + \frac{g}{4}\left[ \cos\varphi_{12}(t) (x_{1}x_{2} + y_{1}y_{2}) + \sin\varphi_{12}(t) (x_{1}y_{2} - y_{1}x_{2}) \right],
\end{equation}
where we defined $\varphi_{12}(t) = (\omega_{1}-\omega_{2})t + \phi_{1} - \phi_{2}$. Next we apply the rotation
\begin{equation}
U_{3} = \exp\left[ \frac{i}{2}(\xi_{1} y_{1} + \xi_{2} y_{2}) \right],
\end{equation}
with $\tan\xi_{k} = \delta_{k}/\Omega_{k}$. The resulting Hamiltonian is
\begin{eqnarray}
H_{3} &=& \frac{1}{2}\left(\frac{\Omega_{1}}{\cos\xi_{1}}x_{1} + \frac{\Omega_{2}}{\cos\xi_{2}}x_{2}\right) + \frac{g}{4} \{ \cos\varphi_{12}(t)[(x_{1}\cos\xi_{1}-z_{1}\sin\xi_{1})(x_{2}\cos\xi_{2}-z_{2}\sin\xi_{2}) \nonumber \\
&+& y_{1}y_{2}] + \sin\varphi_{12}(t)[(x_{1}\cos\xi_{1}-z_{1}\sin\xi_{1})y_{2} - y_{1}(x_{2}\cos\xi_{2}-z_{2}\sin\xi_{2}) ] \},
\end{eqnarray}
where we have used $\delta_{k}\cos\xi_{k}-\Omega_{k}\sin\xi_{k} = 0$ and $\delta_{k}\sin\xi_{k}+\Omega_{k}\cos\xi_{k} = \Omega_{k}/\cos\xi_{k}$. 
The last transformation is given by
\begin{equation}
U_{4} = \exp\left[ -\frac{it}{2}(\eta_{1}x_{1} + \eta_{2}x_{2})\right],
\end{equation}
where $\eta_{k} = \sqrt{\delta_{k}^{2} + \Omega_{k}^{2}}$, such that $\eta_{k} = \Omega_{k}/\cos\xi_{k} = \delta_{k}/\sin\xi_{k}$. This takes our Hamiltonian into the quad frame (QF),
\begin{eqnarray}
H_{4} &=& \frac{g}{4} \{ \cos\varphi_{12}(t) [ x_{1}x_{2}\cos\xi_{1}\cos\xi_{2} - x_{1}\cos\xi_{1}\sin\xi_{2}(z_{2}\cos\eta_{2}t + y_{2}\sin\eta_{2}t) \nonumber \\
&-& (z_{1}\cos\eta_{1}t + y_{1}\sin\eta_{1}t)x_{2}\sin\xi_{1}\cos\xi_{2} + (z_{1}\cos\eta_{1}t + y_{1}\sin\eta_{1}t)(z_{2}\cos\eta_{2}t + y_{2}\sin\eta_{2}t)\sin\xi_{1}\sin\xi_{2} \nonumber \\
&+& (y_{1}\cos\eta_{1}t - z_{1}\sin\eta_{1}t)(y_{2}\cos\eta_{2}t - z_{2}\sin\eta_{2}t) ] + \sin\varphi_{12}(t) [ x_{1}\cos\xi_{1}(y_{2}\cos\eta_{2}t - z_{2}\sin\eta_{2}t) \nonumber \\
&-& (z_{1}\cos\eta_{1}t + y_{1}\sin\eta_{1}t)(y_{2}\cos\eta_{2}t - z_{2}\sin\eta_{2}t)\sin\xi_{1} - (y_{1}\cos\eta_{1}t - z_{1}\sin\eta_{1}t)x_{2}\cos\xi_{2} \nonumber \\
&+& \sin\xi_{2}(y_{1}\cos\eta_{1}t - z_{1}\sin\eta_{1}t)(z_{2}\cos\eta_{2}t + y_{2}\sin\eta_{2}t)] \}.
\end{eqnarray}

Now, we consider the scenario in which we drive the first qubit at the resonance frequency of the second qubit by imposing that $\omega_1 = \omega_{2}^{q}$, while the second qubit is not driven, i.e. $\Omega_{2}=0$, $\eta_{2}=0$, $\delta_{2}=0$, $\omega_{2} = \omega_{2}^{q}$, $\xi_{2}=0$, $\phi_{2}=0$ which implies $\varphi_{12}(t) = \phi_{1}$. The resulting Hamiltonian is
\begin{eqnarray}
\nonumber H_{4} &=& \frac{g}{4} \{ \cos\phi_{1} [ x_{1}x_{2}\cos\xi_{1} - (z_{1}\cos\eta_{1}t + y_{1}\sin\eta_{1}t)x_{2}\sin\xi_{1} + (y_{1}\cos\eta_{1}t - z_{1}\sin\eta_{1}t)y_{2} ] \\
&+& \sin\phi_{1} [ x_{1}y_{2}\cos\xi_{1} - (z_{1}\cos\eta_{1}t + y_{1}\sin\eta_{1}t)y_{2}\sin\xi_{1} - (y_{1}\cos\eta_{1}t - z_{1}\sin\eta_{1}t)x_{2} \},
\end{eqnarray}
where we see that static terms have developed from the slowly rotating terms we kept in the RWA, since with the cross-resonant driving $\omega_{1}-\omega_{2}=\omega_{2}^{q}-\omega_{2}=0$. Finally, we perform a second RWA by dropping any term proportional to $e^{\pm i\eta_{1}t}$, and keep only the static terms. Additionally, we consider the weak-driving regime ($\Omega_{1}/\delta_{1}\ll 1$), which simplifies $\cos\xi_{1}\approx \Omega_{1}/\delta_{1}$. Then, we arrive at the Hamiltonian
\begin{equation}
H_{4} = \frac{g\Omega_{1}}{4\delta_{1}} (x_{1}x_{2}\cos\phi_{1}+x_{1}y_{2}\sin\phi_{1}),
\end{equation}
presented in Eq.~\ref{eq:H_QF}. The validity of this approximation relies on $g/\eta_{1}\approx g/\delta_{1}\ll 1$, which is enforced in the weak-coupling regime. See that the remaining terms after this second RWA are those we kept as slow-rotating after the first RWA, and the terms neglected in this case oscillate with $\delta_{1}=\omega_{1}^{q}-\omega_{1}=\omega_{1}^{q}-\omega_{2}^{q}$.

\subsection{N qubit case}
\label{sec:N_qubit_details}
We start with the N-qubit Hamiltonian in the laboratory frame, given by Eq.~\ref{eq:HamilNq} in the main text. We can move to the QF by applying the following transformations
\begin{equation}
U_{12} = \exp\left[ -\frac{i}{2}\sum_{k=1}^{N}(\omega_{k}t + \phi_{k})z_{k}\right], \hspace{0.3cm}
U_{3} = \exp\left[ \frac{i}{2}\sum_{k=1}^{N}\xi_{k} y_{k} \right] \hspace{0.3cm} \mbox{and} \hspace{0.3cm}
U_{4} = \exp\left[ -\frac{it}{2}\sum_{k=1}^{N}\eta_{k} x_{k} \right].
\end{equation}
Now, as stated in the main text, we drive all qubits at the resonance frequency of their neighbour to the right (except for the last one when applicable). This implies that $\omega_{k}=\omega_{k+1}^{q}$, $\varphi_{k}(t) = \delta_{k+1}t + \phi_{k} - \phi_{k+1}$ and, in the weak-driving regime $\Omega_{k} \ll \delta_{k}$, $\eta_{k}\approx \delta_{k}$. This results in
\begin{eqnarray}\label{H_QF_o}
H_{4} &=& \frac{1}{4} \sum_{k=1}^{N-1} g_{k} \bigg\{ \cos(\delta_{k}t + \phi_{k}-\phi_{k+1}) (y_{k}y_{k+1}+z_{k}z_{k+1}) + \sin(\delta_{k}t + \phi_{k}-\phi_{k+1}) (y_{k}z_{k+1}-z_{k}y_{k+1}) \nonumber\\
&+& \frac{\Omega_{k}}{\delta_{k}} \Big[ \sin(\phi_{k}-\phi_{k+1}) x_{k}y_{k+1} - \cos(\phi_{k}-\phi_{k+1}) x_{k}z_{k+1} \Big] - \frac{\Omega_{k+1}}{\delta_{k+1}} \Big[ \sin[(\delta_{k}+\delta_{k+1})t + \phi_{k}-\phi_{k+1}] y_{k}x_{k+1} \nonumber \\
&+& \cos[(\delta_{k}+\delta_{k+1})t + \phi_{k}-\phi_{k+1}] z_{k}x_{k+1} \Big] \bigg\}.
\end{eqnarray}
The next step is to perform the RWA by neglecting all fast oscillating terms, with frequencies $\delta_k$ and $\delta_k+ \delta_{k+1}$, while keeping the static ones. The resulting Hamiltonian, in the QF, is given by
\begin{equation}
H_{4} = \sum_{k=1}^{N-1}\frac{g_{k}\Omega_{k}}{4\delta_{k}}x_{k}(y_{k+1}\sin(\phi_{k}-\phi_{k+1}) - z_{k+1}\cos(\phi_{k}-\phi_{k+1})),
\end{equation}
as appears in Eq.~\ref{H_eff_QF_complete}.

\section{Unitary transformation to the Quad Frame}
\label{appB}
In order to perform a quantum simulation on the QF, we need to translate the state of our circuit to this frame. Then, considering a simulation scenario in any IBM superconducting chip, we want to find a simple expression for the combination of rotations we need to apply in order to move from IBM's frame into the QF. For that, we will expand the product 
\begin{equation}
U^{\dagger}_{\text{IBM}}U_{12}U_{3}U_{4} = e^{\frac{it}{2}\sum_{k=1}^{N-1}\omega^{q}_{k}z_{k}} e^{-\frac{it}{2}\sum_{k=1}^{N-1}\omega_{k}z_{k}} e^{\frac{i}{2}\sum_{k=1}^{N-1}\xi_{k} y_{k}} e^{-\frac{it}{2}\sum_{k=1}^{N-1}\eta_{k} x_{k}},
\end{equation}
having set $\phi_{k}=\phi=0$. See that the first two exponentials can be combined, such that
\begin{equation}
U^{\dagger}_{\text{IBM}}U_{12}U_{3}U_{4} =  e^{\frac{it}{2}\sum_{k=1}^{N-1}\delta_{k}z_{k}} e^{\frac{i}{2}\sum_{k=1}^{N-1}\xi_{k} y_{k}} e^{-\frac{it}{2}\sum_{k=1}^{N-1}\eta_{k} x_{k}},
\end{equation}
where $\delta_{k} = \omega_{k}^{q}-\omega_{k}$. Now, it is satisfied that
\begin{equation}
 \exp\left[ i\sum_{k=1}^{N}\theta_{k} \sigma_{k} \right] = \prod_{k=1}^{N} e^{i\theta_{k}\sigma_{k}}
\end{equation}
for $\sigma = x$, $y$, or $z$. This means that we can write
\begin{equation}
U^{\dagger}_{\text{IBM}}U_{12}U_{3}U_{4} =  \prod_{k=1}^{N-1} e^{\frac{it}{2}\delta_{k}z_{k}} e^{\frac{i}{2}\xi_{k}y_{k}} e^{-\frac{it}{2}\eta_{k}x_{k}},
\end{equation}
and we can use the Euler form for Pauli matrices,
\begin{equation}
e^{ i \theta \sigma} = \cos\theta \, \mathbb{1} + i\sin\theta \, \sigma,
\end{equation}
to express these rotations as 
\begin{equation}
U^{\dagger}_{\text{IBM}}U_{12}U_{3}U_{4} =  \prod_{k=1}^{N-1} (\cos\frac{\delta_{k}t}{2} \mathbb{1}_{k} + i\sin\frac{\delta_{k}t}{2} z_{k}) (\cos\frac{\xi_{k}}{2} \mathbb{1}_{k} + i\sin\frac{\xi_{k}}{2} y_{k}) (\cos\frac{\eta_{k}t}{2} \mathbb{1}_{k} - i\sin\frac{\eta_{k}t}{2} x_{k}).
\end{equation}
Recall that, working in the regime $\Omega \ll \delta$, we had approximated $\eta \approx \delta$, $\sin\xi \approx 1$, and $\cos\xi \approx \Omega/\delta$. Knowing that $\sin\theta/2 = \sqrt{(1-\cos\theta)/2}$ and $\cos\theta/2 = \sqrt{(1+\cos\theta)/2}$, we can simplify
\begin{equation}
\cos\frac{\xi_{k}t}{2} \mathbb{1}_{k} + i\sin\frac{\xi_{k}t}{2} y_{k} \approx \frac{1}{\sqrt{2}} \left( \mathbb{1}_{k} + i y_{k} + \frac{\Omega_{k}}{2\delta_{k}}(\mathbb{1}_{k} - i y_{k})\right),
\end{equation}
where we have used $\sqrt{1\pm x}\approx 1\pm x/2$ for small $x$. In this expansion, we eventually find
\begin{equation}
U^{\dagger}_{\text{IBM}}U_{12}U_{3}U_{4} =  \prod_{k=1}^{N-1} \frac{1}{\sqrt{2}}\bigg[ \mathbb{1}_{k} + i y_{k}+ \frac{\Omega_{k}}{2\delta_{k}} ((\mathbb{1}_{k} - i y_{k}) \cos\delta_{k}t  + i(z_{k}-x_{k}) \sin\delta_{k}t ) \bigg],
\end{equation}
which we will denote by $U_{\text{QF}}$. Let us check the unitarity of this operator by computing
\begin{equation}
U_{\text{QF}}U_{\text{QF}}^{\dagger} = \mathbb{1} + \mathcal{O}\left(\frac{\Omega^{2}}{\delta^{2}}\right).
\end{equation}
The previous calculations were set in the weak-driving regime ($\Omega/\delta \ll 1$), considering terms up to first order in $\Omega/\delta$ and neglecting higher orders. This is consistent with the approximations we have made here, and thus the unitarity of $U_{\text{QF}}$ relies on these approximations. 


\section{Synthesis errors}
\label{app:synthesiserror}
In this appendix, we want to show the synthesis errors corresponding to the toggled Hamiltonians. For the $XY$ model, the original Hamiltonian is
\begin{eqnarray}
H^{org}_{XY} &=& \frac{g}{4} \sum_{k=1}^{N-1} \bigg\{ - \frac{\Omega}{\delta}(y_{k}y_{k+1} + x_{k}x_{k+1}) \\
\nonumber &+& (x_{k}y_{k+1}+y_{k}x_{k+1}-2z_{k}z_{k+1})\cos\delta t + [z_{k}(y_{k+1}-x_{k+1}) + (x_{k}-y_{k})z_{k+1}]\sin\delta t \\
\nonumber &+& \frac{\Omega}{\delta} [z_{k}(y_{k+1}-x_{k+1})\sin 2\delta t - (y_{k}y_{k+1} + x_{k}x_{k+1})\cos 2\delta t] \bigg\}.
\end{eqnarray}
Then, the difference between original and effective Hamiltonians,
\begin{eqnarray}
\Delta H_{XY} &=& \frac{g}{4} \sum_{k=1}^{N-1} \bigg\{ (x_{k}y_{k+1}+y_{k}x_{k+1}-2z_{k}z_{k+1})\cos\delta t + [z_{k}(y_{k+1}-x_{k+1}) + (x_{k}-y_{k})z_{k+1}]\sin\delta t \\
\nonumber &+& \frac{\Omega}{\delta} [z_{k}(y_{k+1}-x_{k+1})\sin 2\delta t - (y_{k}y_{k+1} + x_{k}x_{k+1})\cos2\delta t] \bigg\},
\end{eqnarray}
constitutes the error we want to estimate. We find the Frobenius norm is given by
\begin{equation}
|| \Delta H_{XY} ||_{F} = \frac{g}{2}\sqrt{N-1} .
\end{equation}
On the other hand, the original $ZZ$ toggled Hamiltonian is
\begin{eqnarray}
\nonumber H^{org}_{ZZ} &=& \frac{g}{4} \sum_{k=1}^{N-1} \bigg\{ \left[ z_{k}\frac{\Omega}{\delta} - x_{k}\cos\delta t + y_{k}\sin\delta t \right] z_{k+1} + (y_{k}\cos\delta t + x_{k}\sin\delta t)y_{k+1} \\
\nonumber &+& \cos\varphi_{k}(t) \left[  z_{k}(z_{k+1}\frac{\Omega}{\delta} -x_{k+1}\cos\delta t + y_{k+1}\sin\delta t) + y_{k}(y_{k+1}\cos\delta t + x_{k+1}\sin\delta t) \right] \\
&+& \sin\varphi_{k}(t) \left[ -z_{k}(y_{k+1}\cos\delta t + x_{k+1}\sin\delta t) + y_{k}(z_{k+1}\frac{\Omega}{\delta} - x_{k+1}\cos\delta t + y_{k+1}\sin\delta t) \right] \bigg\},
\end{eqnarray}
so that the difference is
\begin{eqnarray}
\nonumber \Delta H_{ZZ} &=& \frac{g}{4} \sum_{k=1}^{N-1} \bigg\{ (- x_{k}\cos\delta t + y_{k}\sin\delta t ) z_{k+1} + (y_{k}\cos\delta t + x_{k}\sin\delta t)y_{k+1} \\
\nonumber &+& \cos\varphi_{k}(t) \left[  z_{k}(z_{k+1}\frac{\Omega}{\delta} -x_{k+1}\cos\delta t + y_{k+1}\sin\delta t) + y_{k}(y_{k+1}\cos\delta t + x_{k+1}\sin\delta t) \right] \\
&+& \sin\varphi_{k}(t) \left[ -z_{k}(y_{k+1}\cos\delta t + x_{k+1}\sin\delta t) + y_{k}(z_{k+1}\frac{\Omega}{\delta} - x_{k+1}\cos\delta t + y_{k+1}\sin\delta t) \right] \bigg\}.
\end{eqnarray}
The Frobenius norm is then given by
\begin{equation}
|| \Delta H_{ZZ} ||_{F} = \frac{g}{2\sqrt{2}}\sqrt{N-1}\sqrt{2+\cos\delta t\cos(\varphi_{k}(t)-\delta t)+\frac{\Omega}{\delta}\sin\delta t\sin\varphi_{k}(t)}.
\end{equation}

\section{2D $XY$ model}
\label{app:norm}
In this appendix, we describe the transformation from a spin lattice to a string of fermions with 2-site hopping, which allows us to estimate more accurately the Trotter error associated to the simulation of the $XY$ model in 2 dimensions. This error will be given by the commutator $[H_{I},H_{II}]$, split into $A =  [H_{I}^{yy},H_{II}^{xx}]$ and $B = [H_{I}^{xx},H_{II}^{yy}]$. Both $A$ and $B$ include 8 terms, each one having the form $x$-$z$-$y$ at three different vertices, as we can see in Fig.~\ref{fig7}. Assembling pairs of these terms, joining $x_{i}z_{i+1}y_{i+2}$ with $y_{i+2}z_{i+3}x_{i+4}$ or $y_{i}z_{i+1}x_{i+2}$ with $x_{i+2}z_{i+3}y_{i+4}$, either in $A$ or in $B$, we can construct diagonal strings in the lattice, which we also represent in Fig.
~\ref{fig7}. These diagonals can then be thought of as 1-d strings, and given this outline, we can apply a Jordan-Wigner transformation and introduce Majorana fermion operators. Let us take a couple of terms (1 \& 7) in $B$ to illustrate this issue:
 \begin{equation}
 \sum_{i,j} (x_{2i,2j-1}z_{2i-1,2j-1}y_{2i-1,2j} + x_{2i,2j+1}z_{2i,2j}y_{2i+1,2j}),
 \end{equation}
can be turned into a string as
 \begin{equation}
 \sum_{j=1}^{N^{2}/4} (x_{4j-3}z_{4j-2}y_{4j-1} + y_{4j-1}z_{4j}x_{4j+1}).
 \end{equation}

\begin{figure}[t]
{\includegraphics[width=0.75 \textwidth]{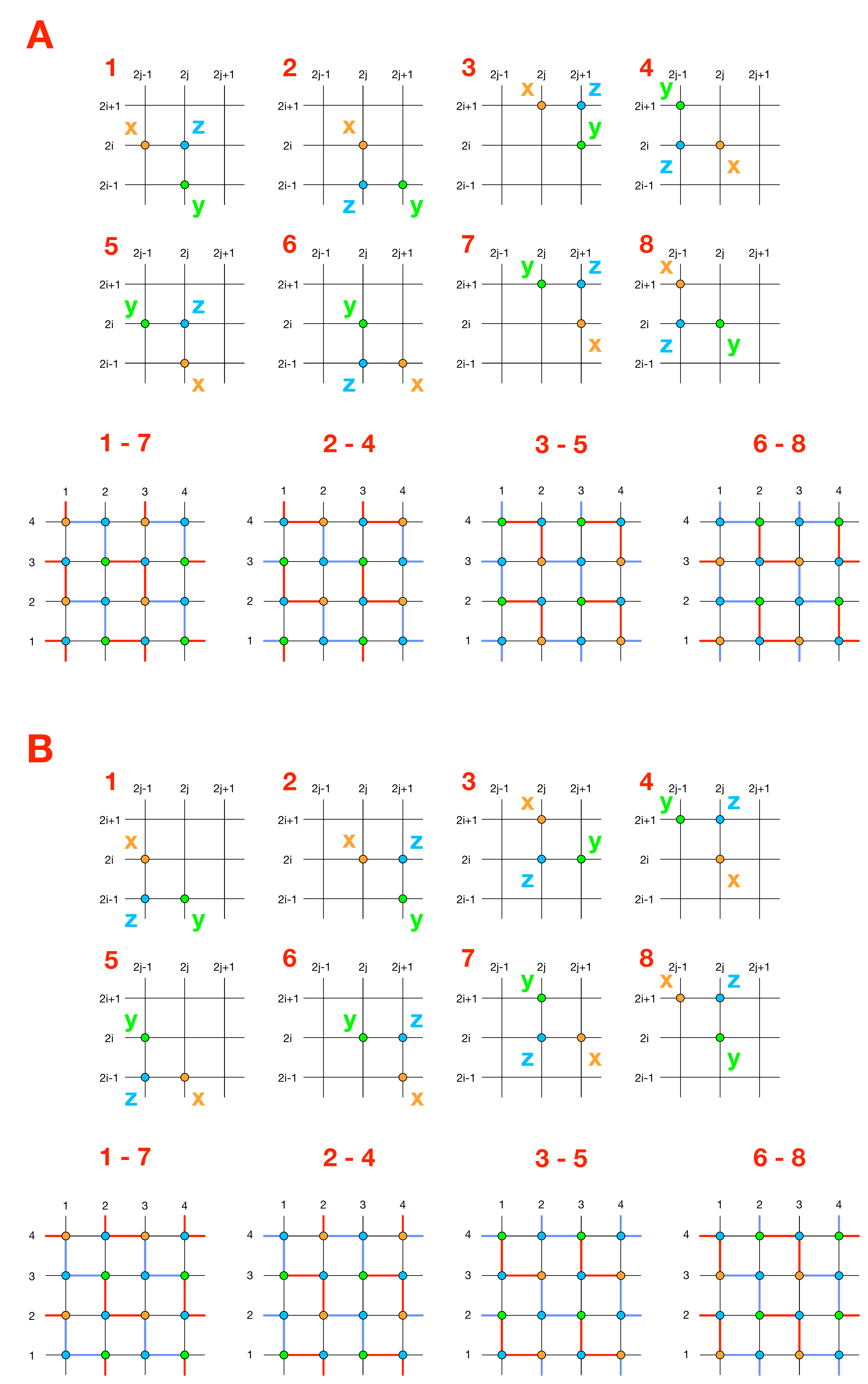}}
\caption{Representation of the spin triplets, grouped in $A$ and $B$, which constitute Table~\ref{table1}. These triplets, inside either $A$ or $B$, can be grouped in pairs, which can be used to tile the entire lattice with staircase patterns. These tilings form strings with periodic boundary conditions along the diagonals of the lattice, which we transform from spins to fermions in order to estimate $||A||+||B||$, the Trotter error associated to the digital-analog simulation of the $XY$ model in 2D.}
\label{fig7}
\end{figure}

The Jordan-Wigner transformation, followed by defining Majorana operators $\gamma^{(1)}_{i} = c_{i}+c^{\dagger}_{i}$ and $\gamma^{(2)}_{i} = -i(c^{\dagger}_{i}-c_{i})$, transforms spins as
\begin{equation}
x_{j}z_{j+1}y_{j+2} = -i \gamma_{j}^{(2)}\gamma_{j+2}^{(2)}
\end{equation}
and
\begin{equation}
y_{j}z_{j+1}x_{j+2} = i \gamma_{j}^{(1)}\gamma_{j+2}^{(1)}.
\end{equation}
These transformations lead to the Hamiltonian
\begin{equation}
-i \sum_{j=1}^{N^{2}/4} \left(\gamma^{(2)}_{4j-3}\gamma^{(2)}_{4j-1}-\gamma^{(1)}_{4j-1}\gamma^{(1)}_{4j+1}\right).
 \end{equation}
 The Fourier transform of Majorana operators is given by
\begin{eqnarray}
\nonumber \gamma^{(1)}_{j} &=& \frac{1}{\sqrt{N}} \sum_{k}\left(\gamma^{(1)}_{k}\cos kj + \gamma^{(2)}_{k}\sin kj \right), \\
\gamma^{(2)}_{j} &=& \frac{1}{\sqrt{N}} \sum_{k}\left(\gamma^{(2)}_{k}\cos kj - \gamma^{(1)}_{k}\sin kj \right),
\end{eqnarray}
together with
\begin{eqnarray}
\nonumber \sum_{j}\gamma^{(1)}_{j}\gamma^{(1)}_{j+2} &=& \frac{1}{2}\sum_{k}\Big[ \cos2k \left(\gamma^{(1)}_{k}\gamma^{(1)}_{k}+\gamma^{(2)}_{k}\gamma^{(2)}_{k}+\gamma^{(1)}_{k}\gamma^{(1)}_{-k}-\gamma^{(2)}_{k}\gamma^{(2)}_{-k} \right) \\
\nonumber &+& \sin2k \left(\gamma^{(1)}_{k}\gamma^{(2)}_{k}-\gamma^{(2)}_{k}\gamma^{(1)}_{k}-\gamma^{(1)}_{k}\gamma^{(2)}_{-k}-\gamma^{(2)}_{k}\gamma^{(1)}_{-k} \right)  \Big] \\
\nonumber \sum_{j}\gamma^{(2)}_{j}\gamma^{(2)}_{j+2} &=& \frac{1}{2}\sum_{k}\Big[ \cos2k \left(\gamma^{(1)}_{k}\gamma^{(1)}_{k}+\gamma^{(2)}_{k}\gamma^{(2)}_{k}-\gamma^{(1)}_{k}\gamma^{(1)}_{-k}+\gamma^{(2)}_{k}\gamma^{(2)}_{-k} \right) \\
&+& \sin2k \left(\gamma^{(1)}_{k}\gamma^{(2)}_{k}-\gamma^{(2)}_{k}\gamma^{(1)}_{k}+\gamma^{(1)}_{k}\gamma^{(2)}_{-k}+\gamma^{(2)}_{k}\gamma^{(1)}_{-k} \right)  \Big],
\end{eqnarray}
which leads to the expression of a string in Fourier space,
\begin{equation}
i \sum_{k} \left[ \cos 2k \left(\gamma^{(1)}_{k}\gamma^{(1)}_{-k}-\gamma^{(2)}_{k}\gamma^{(2)}_{-k} \right) - \sin 2k \left(\gamma^{(1)}_{k}\gamma^{(2)}_{-k}+\gamma^{(2)}_{k}\gamma^{(1)}_{-k} \right)\right].
\end{equation}
 The resulting matrix is block-diagonal, and it has two types of blocks. The first block, $\Gamma_{1}$, contains the elements $\cos 2k$ for $k\in[-\pi,\pi]$, and the second block, $\Gamma_{2}$, contains the elements $\sin 2k$. Both blocks have elements only along the anti-diagonal. Then, the matrix in Fourier space looks like
 \begin{equation}
 \begin{pmatrix} \Gamma_{1} & -\Gamma_{2} \\ -\Gamma_{2} & -\Gamma_{1} \end{pmatrix}.
 \end{equation}
We can write the eigenvalue problem as 
\begin{equation}
\det \left[  \begin{pmatrix} \Gamma_{1} & -\Gamma_{2} \\ -\Gamma_{2} & -\Gamma_{1} \end{pmatrix} -  \begin{pmatrix} \mathbb{1}\lambda_{+} & 0 \\ 0 & \mathbb{1}\lambda_{-} \end{pmatrix}\right] = 0,
\end{equation}
for $\lambda_{+}=\lambda$ and $\lambda_{-}=-\lambda$. Given that the determinant of this matrix is $-\det(\Gamma_{1}+\Gamma_{2}\Gamma_{1}^{-1}\Gamma_{2})\det\Gamma_{1} = 1$, we have a hint that $\lambda = \pm 1$, and in fact 
\begin{equation}
\det \left[  \begin{pmatrix} \Gamma_{1} & -\Gamma_{2} \\ -\Gamma_{2} & -\Gamma_{1} \end{pmatrix} -  \begin{pmatrix} \mathbb{1}\lambda & 0 \\ 0 & -\mathbb{1}\lambda \end{pmatrix}\right] = (1-\lambda^{2})^{N^{2}/2} = 0
\end{equation}
As we can see, the eigenvalues of this matrix are $\lambda_{j}=\pm 1$. Then, the spectral norm of this pair of elements in the commutator is equal to one, for any lattice size. Since there are 4 total pairs in $B$, we will get that the norm can be estimated as $||B|| \leq 8 J^{2}$, taking into account the $2i$ factor from Pauli commutation relations, and the $J^{2}$ from the analog Hamiltonian. Given that the norm of $A$ can be estimated in the same way, we have
\begin{equation}
||[H_{I},H_{II}]|| = || - A+B || \leq ||A|| + ||B|| \leq 16 J^{2},
\end{equation}
where we have eliminated the dependence on the system size. 


\end{widetext}

\end{document}